\documentclass[aps, pre, twocolumn, groupedaddress]{revtex4-2}

\usepackage{graphicx}
\usepackage{subcaption}
\usepackage{xcolor}
\usepackage{mathrsfs}
\usepackage{amsmath,amssymb}
\usepackage{amsthm}
 \usepackage[colorlinks=true, linkcolor=blue, citecolor=blue, urlcolor=blue]{hyperref} 

\theoremstyle{remark}
\newtheorem{remark}{Remark} 
\newtheorem*{remark*}{Remark} 

\begin{document}
\title{On the Equivariant Learning of the $Q$-tensor Order Parameter}
\author{Julia Navarro ($\dagger$), Mark Wilkinson ($\ddagger$)}
\affiliation{$\dagger$ Nottingham Trent University, UK, $\ddagger$ Berea College, USA}
\email{($\dagger$) julia.navarro2019@my.ntu.ac.uk (Corresponding Author), \\($\ddagger$) wilkinsonm@berea.edu}


\begin{abstract}
We construct and evaluate group-equivariant neural networks for the prediction of the two-dimensional $Q$-tensor order parameter of nematic liquid crystals from synthetically generated microscopic textures. Seven architectures, equivariant to cyclic groups $C_k$ of order $k$ for $k=4,\,8,\,16,\,32,\,64,\,128,\, 256$, are built using a combination of weight-sharing constraints, equivariant activations and regularization techniques. To do this, we construct rotation-like permutation matrix groups with elements $\varrho_{C_k}(g)$ that act on row-wise vectorized images, thereby approximating a $\frac{2\pi}{k}$ rotation of the circular subdomain on square images.
We show that all seven equivariant models satisfy the $Q$-tensor equivariance constraint to within single-precision floating point accuracy. Comparing against approximate parameter-matched non-equivariant benchmarks, with and without data augmentation, we find that the equivariant models consistently achieve lower errors and generalize more robustly to unseen defect configurations. Performance increases with group order, suggesting that the incorporation of finer rotational symmetry leads to lower errors.   

\end{abstract}

\maketitle



\subsection{Introduction}


Machine Learning has increasingly been used in material science and soft matter physics to overcome computational bottlenecks of traditional analytical models and simulation-based methods, such as in molecular dynamics \cite{Stavrogiannis2026}. Whilst early applications to problems involving liquid crystals focused on simple tasks, such as predicting clearing temperatures from molecular structures \cite{kranz1996prediction}, more recent applications go beyond simple regression and are used to interpret Polarised Optical Microscopy (POM) textures. Several works explore temperature, phase transitions, and parameter prediction from POM images using various model types: estimating physical parameters using the $k$-nearest neighbour algorithm in \cite{Sigaki_2019}, prediction of liquid crystal properties and parameters using Convolutional Neural Networks in \cite{Sigaki_2020}, hybrid CNN-Recurrent Neural Network for parameter prediction in \cite{Colen2021}, classification of textures using CNNs in \cite{Dierking02Dec2022}, classification of liquid crystal phases with CNN and Inception models in \cite{Dierking7June2023}, liquid crystal phase classification with CNNs in \cite{Betts2023}, CNNs and Inception models used to identify transition temperatures and phases in \cite{Dierking3Feb2023}, Neural Network (NN) prediction of elastic constants in \cite{Zaplotnik2023}, and classification using local binary patterns algorithm and CNNs of phases in \cite{Osiecka-Drewniak26012024}. More recently, the authors in \cite{Terroa2025} trained CNNs on simulated POM textures to predict free energy,
cholesteric pitch, and strength of applied electric fields.

Amongst other machine learning methods, Random Forests (RFs), have been used for the prediction of order parameters \cite{Inokuchi2020}. In \cite{Takahashi2022} , the authors formalised their RF approach in the learning package MALIO where they were able to distinguish between different LC phases and select candidate local order parameters used to observe the phase transition. Another type of ML method investigated are Support Vector Machines (SVMs) by \cite{Liu_2019}, used for multi-class phase classification, alongside being able to determine the presence of a transition. Furthermore, \cite{Shi2024} developed a NN-based tensor model mapping the $Q$-tensor to Maier-Saupe bulk energies, surpassing Landau-de Gennes in capturing phase transitions and defect configurations. The authors in \cite{Beyerle2025} showed that generative score-based models (Thermodynamic Maps) can infer isotropic-nematic phase behaviour from minimal training data, outperforming Principal Component Analysis (PCA) and Variational Auto-encoders (VAEs). All of the above-mentioned works demonstrate the relevance of machine learning to material science.

In this work, we contribute to the study of nematic liquid crystals by constructing neural networks which can predict the de Gennes $Q$-tensor order parameter from microscopic data with high accuracy, highlighting in the process the practical advantages to working with \emph{group-equivariant} networks which inherit the equivariance property of the underlying order parameter.
\section{Background}


We give a brief background on Liquid Crystals, the $Q$-tensor Order parameter, and Equivariant Neural Networks (ENNs). We also present to the reader how neural networks can learn order parameters from molecular data in the presence of symmetry and equivariance.

\subsection{Liquid Crystal Order Parameters}


Liquid crystals (LCs) are a distinct phase of matter whose physical behaviour lies between that of crystalline solids and isotropic liquids \cite{10.1093/oso/9780198520245.001.0001}. They flow like fluids whilst exhibiting anisotropic (locally ordered) optical properties characteristic of crystals. First identified in cholesterol derivatives by Reinitzer in 1888 \cite{Reinitzer1888} and later termed by Lehmann, this phase has become foundational to both the theoretical study of complex fluids and technological applications, most notably in display technology (e.g., LCDs).\\

\begin{figure}[h] 
\centering
\includegraphics[width=1.0\columnwidth]{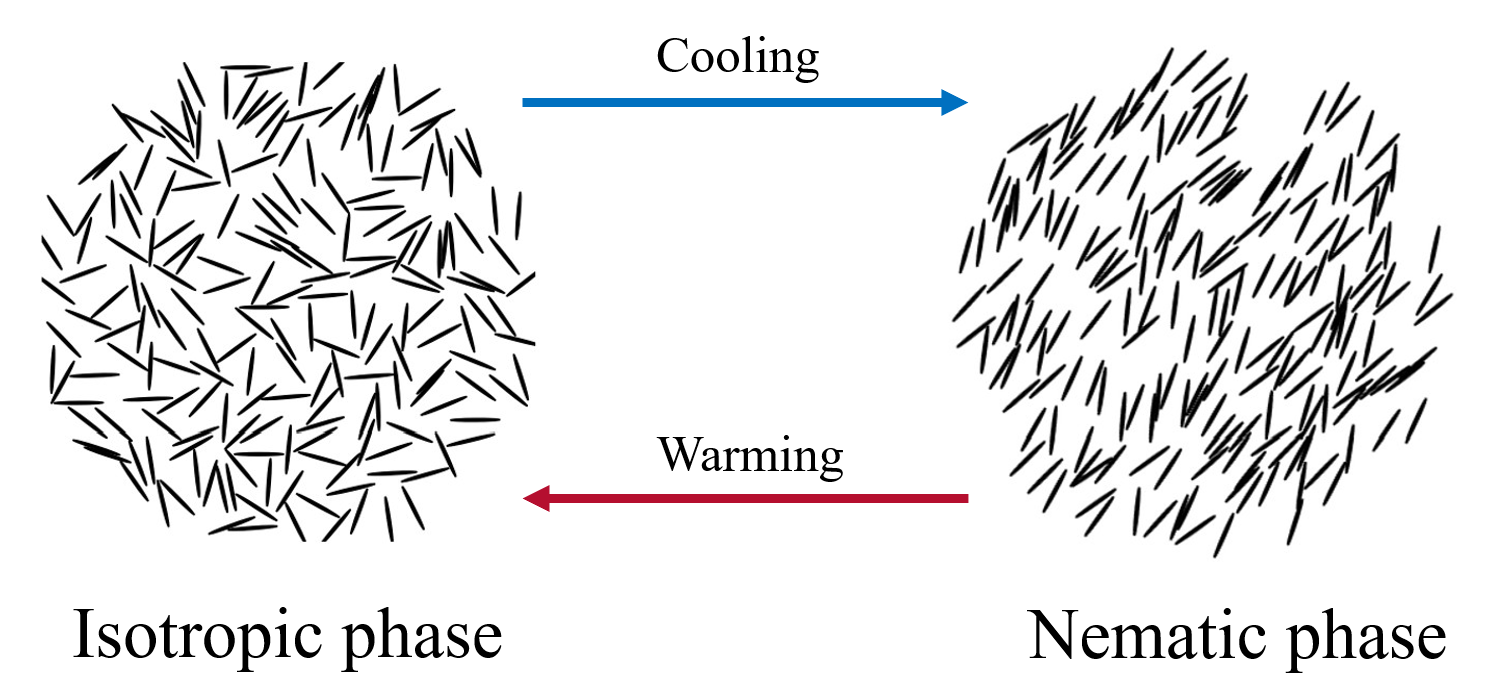}
\caption{\scriptsize{Schematic illustration of the isotropic-to-nematic phase transition in a liquid crystal. Upon cooling below the transition temperature, the rod-like molecules transition from a disordered isotropic phase (random orientations) to a nematic phase with long-range orientational order along a preferred direction (e.g. the director). No positional ordering is present in nematics.}}
\label{Isotropic to Nematic Diagram}
\end{figure}

Liquid crystal phases are commonly classified according to the physical mechanisms driving their phase transitions. In thermotropic liquid crystals (see Figure \ref{Isotropic to Nematic Diagram}), phase changes are induced by variations in temperature, whereas lyotropic LCs admit a transition based on embedded  amphiphile concentration. Among these, the nematic phase is the most extensively studied in the literature, and is the phase on which we shall focus in this article. Nematic LCs are characterized by long-range orientational order but lack the long-range positional order present in more structured phases such as smectics.
Different order parameters exist to study such materials, such as the scalar order parameter $S$ \cite{mottram2014introductionqtensortheory}, the director $\hat{n}$ \cite{mottram2014introductionqtensortheory}, or probability distribution functions as found in Maier-Saupe theory \cite{Dissipative_Ordered_Fluids}. In this article, we shall work with the well studied case of two-dimensional nematic textures, which can be thought of as thin-film or planar arrangements of nematic molecular systems: see the work of Zarnescu et al. \cite{doi:10.1142/S0218202515500396, frankyandzarnescu} and others \cite{https://doi.org/10.1111/sapm.12161}. The particular order parameter we shall use to describe the local structure in these systems is the $Q$-tensor order parameter of de Gennes. In this two-dimensional case, the $Q$-tensor order parameter $Q\in\mathbb{R}^{2\times 2}$ is defined in the following way:
\begin{equation}\label{qdefcont}
Q=Q[\rho]:=\int_{\mathbb{S}^{1}}\left(\vec{n}\otimes\vec{n}-\frac{1}{2}\mathbb{I}\right)\rho(\vec{n}),
\end{equation}
where $\rho:\mathbb{S}^{1}\rightarrow [0, 1]$ is a probability density function that admits the antipodal symmetry condition $\rho(-\vec{n})=\rho(\vec{n})$ for all $\vec{n}\in\mathbb{S}^{1}$, modeling head-to-tail symmetry of the underlying nematic molecules. As a consequence of this definition, $Q$ admits the symmetry $Q_{ij}=Q_{ji}$ ($Q^T = Q$), is traceless (tr$(Q) = 0$), and has bounded eigenvalues ($-\frac{1}{2}\leq\lambda(Q)\leq  \frac{1}{2}$): see, for instance,  \cite{ball_cambridge_slides}. In other words, the order parameter $Q\in\mathbb{R}^{2\times 2}$ is necessarily of the form
%
%
%
\begin{equation*}
    Q  =
    \begin{pmatrix}
     Q_{11} & Q_{12} \\
     Q_{12} & -Q_{11}
    \end{pmatrix},
\end{equation*}
where $Q_{11}$ and $Q_{12}$ are two real numbers. For further information on the properties of $Q$ and its definition in terms of molecular orientation, see \cite{10.1093/oso/9780198520245.001.0001}.

As an order parameter, the values of $Q$ give rough statistical information about the local alignment of nematic molecular systems. Notably, values of $Q$ close to the zero matrix $0\in\mathbb{R}^{2\times 2}$ describe the isotropic phase (in which molecules are randomly oriented in space), whereas values of $Q$ for which $0<|Q_{11}|, |Q_{12}|$ describe the nematic phase (in which molecular orientations are locally correlated).

From the point of view of machine learning, the definition \eqref{qdefcont} of $Q$ gives rise to a mapping, whose inputs are probability distribution functions $\rho$ and whose outputs are traceless and symmetric matrices with bounded eigenvalues. As such, one might try to learn this mapping using supervised learning with a suitable training set $\mathscr{S}:=\{(\rho_{k}, Q_{k})\}_{k}\subset [0, 1]^{\mathbb{S}^{1}}\times \mathbb{R}^{2\times 2}$ and an appropriate network architecture.  


In this article, as we work with explicit molecular texture images such as those appearing in Figure \ref{Isotropic to Nematic Diagram}, it is more appropriate to work with discrete empirical measures $\mu_{N}$ on the unit circle $\mathbb{S}^{1}$ as opposed to continuously-distributed probability distribution functions $\rho$ as appearing in \eqref{qdefcont}. In this way, we can view the $Q$-tensor order parameter as a mapping from discrete measures $\mu_{N}$ of the form
\begin{equation*}
\mu_{N}:=\frac{1}{2N}\sum_{i=1}^{N}\left(\delta_{\vec{n}(\theta_{i})}+\delta_{-\vec{n}(\theta_{i})}\right)    
\end{equation*}
to traceless and symmetric $2\times 2$ matrices with bounded eigenvalues through the formula
\begin{equation}\label{Eq: Integral form of Q}
\begin{array}{c}
    Q = Q[\mu_N] = \displaystyle\int_{\mathbb{S}^1} \left(\vec{n}\otimes \vec{n} - \frac{1}{2}\mathbb{I}\right)d\mu_N(\vec{n}) \vspace{2mm}\\
    \displaystyle =\frac{1}{2N}\sum_{i=1}^{N}\left(\vec{n}(\theta_{i})\otimes \vec{n}(\theta_{i})-\frac{1}{2}\mathbb{I}\right),
\end{array}
\end{equation}
where $\vec{n}(\theta)$ is the unit vector
\begin{equation*}
\vec{n}(\theta):=\left(
\begin{array}{c}
\cos(\theta) \\
\sin(\theta)
\end{array}
\right)
\end{equation*}
for $\theta\in\mathbb{R}$.
To derive the $Q$-tensor values from nematic texture images such as those in Figure \ref{Isotropic to Nematic Diagram}, we consider each molecular angle $\theta_i$ present in that texture. Given a collection of $N$ nematic molecules, the components of $Q$ are computed by averaging functions of the orientations $\theta_i$ of all individual components: 
\begin{align}\label{Discrete Q}
\begin{split}
  Q_{11} &= \frac{1}{2N} \sum_{i=1}^{N} \left( \cos^2 (\theta_i) - \frac{1}{2} \right) 
= \frac{1}{4N} \sum_{i=1}^{N} \cos(2\theta_i), \\
Q_{12} &= \frac{1}{2N} \sum_{i=1}^{N} \left( \sin (\theta_i) \cos (\theta_i) \right) 
= \frac{1}{4N} \sum_{i=1}^{N} \sin(2\theta_i).  
\end{split}
\end{align}
In this way, every image containing $N$ congruent nematic molecules can be identified with an empirical measure $\mu_{N}$ on $\mathbb{S}^{1}$. Mirroring our comment above in the case of continuously distributed probability density functions, one might hope to learn the $Q$-tensor mapping using supervised learning with a suitable training set $\mathscr{S}':=\{(I_{k}, Q_{k})\}_{k}\subset \mathrm{Space\,of\,Images}\times\mathbb{R}^{2\times 2}$ and an appropriate network architecture. The ``Space of Images'' with which we work in this article is available upon request. Whilst our training images of nematics are synthetic in this study, in that we work with computationally generated packings of elliptical particles, the methods we employ in this work may be useful in the analysis of real-world nematic textures in the laboratory or in industry. 
\begin{remark}
As we identify a discrete measure $\mu_{N}$ with an image containing $N$ elliptical particles with orientation vector $\theta=(\theta_{1}, ..., \theta_{N})$ in all that follows, we shall write the mapping expression $Q[\mu_{N}]$ simply as $Q[\theta]$ throughout, when the underlying vector of angles $\theta$ is understood from the context.
\end{remark}

\subsubsection{Contribution of this Article}
Our work in this article complements, but differs from, the existing work performed in the literature to date. Notably, our study pertains to (simulated) \textit{molecular} systems of nematics. In contrast, many of the pre-existing works in the literature focus on meso- or macroscopic systems of either simulated or experimental liquid crystal materials. Moreover, our work is centered on the novel construction of a network which can predict the $Q$-tensor order parameter from image data with high accuracy.

More recently, and perhaps closely related with the content of this work, Shi et al. \cite{Shi2024} employ neural network-based tensor models, which are in turn supervised by a molecular model, to attain energy precision in modelling nematics at the mesoscopic level which is comparable to molecular ones. Notably, the authors construct a neural network which maps the $Q$-tensor order parameter to the microscopic Maier-Saupe energy. Rather, our work maps images of nematic molecular configurations to the $Q$-tensor order parameter, which might then in turn be fed to models such as those in that study.

\subsection{Equivariance of the $Q$-tensor Mapping}
The Q-tensor mappings \eqref{qdefcont} and \eqref{Eq: Integral form of Q} admit a convenient symmetry property which one would like any neural network trying to learn $Q$ also to inherit. The discrete $Q$-tensor satisfies the following \textbf{equivariance property}:
\begin{equation}\label{Eq: Equiv. Discreet Q}
Q[T_\alpha\theta] = R(\alpha) \, Q[\theta] \, R(\alpha)^T,
\end{equation}
where
$$R(\alpha):= \begin{pmatrix} \cos (\alpha) & -\sin(\alpha) \\ \sin(\alpha) & \cos(\alpha) \end{pmatrix}$$
is the two-dimensional (counter-clockwise) rotation matrix, $\theta = (\theta_1, \hdots, \theta_N)$,
and where $T_\alpha$ denotes the group action of $\mathbb{S}^1$ on $\mathbb{T}^{N}$ defined by $T_\alpha \theta = (\theta_1+\alpha, \hdots, \theta_N+\alpha)$. This means that if all molecule angles $\theta$ are rotated by $\alpha$ counter-clockwise, the resulting $Q$-tensor is conjugated by the counter-clockwise $2\times2$ rotation matrix $R(\alpha)$. In this way, the $Q$-tensor is rotation equivariant. {More on group actions and theory can be found in \cite{rotman1999introduction}, who provides an introduction to the topic.} It is therefore natural that we prescribe that any neural network $\nu:\mathrm{Space\,of\,Images}\rightarrow\mathbb{R}^{2\times 2}$ also admit the same equivariance property, namely
\begin{equation*}
\nu(T_{\alpha} I)=R(\alpha)\nu(I)R(\alpha)^{T},    
\end{equation*}
where $I\in\mathrm{Space\,of\,Images}$ and the group action $T_{\alpha}I$ of $\mathbb{S}^{1}$ acting on the \textrm{Space\,of\,Images} admits a suitable interpretation: see Appendix \ref{Construction of the Rotation-like Permutation Matrix}, as well as our \href{https://github.com/NavarroJulia/On-the-Equivariant-Learning-of-the-Q--Tensor-Order-Parameter}{GitHub} for further information.\\
\begin{remark}
Note that the equivariance condition in (\ref{Eq: Equiv. Discreet Q}) can actually be redefined on the two independent variables of $Q$ as
\begin{align}\label{Eq: Vector form of Q equivariance}
    \vec{q}[{T_\alpha}\theta] = R(2\alpha) \vec{q}[\theta],
\end{align}
where $\vec{q}[\theta]:= (Q_{11}[\theta], Q_{12}[\theta])^T$.
\end{remark}
%





\subsection{Equivariant Neural Networks}



\subsubsection{Literature Review on Equivariant Learning}

Machine learning has become foundational in data-driven scientific research \cite{2015Natur.521..436L}. Since the development of multi-layer feedforward networks and the popularization of the backpropagation algorithm by \cite{Rumelhart1986LearningRB}, a further resurgence of interest in artificial neural networks followed the landmark application of deep convolutional neural networks in the early 2010s, most notably ImageNet in 2012 \cite{10.1145/3065386}. Whilst multilayer feedforward networks are considered universal approximators \cite{HORNIK1989359, Cybenko1989ApproximationBS}, fully connected layers lead to an explosive growth in the number of parameters (e.g., tens or hundreds of thousands of weights even modest image sizes). This motivates the use of weight-sharing and symmetry-exploiting architectures. One such example is the Convolutional Neural Network (CNN) \cite{726791}, which has revolutionalized image processing \cite{10.1145/3065386}. This is due to the weight sharing mechanism and translation equivariance of convolutional networks \cite{726791, cohen2016groupequivariantconvolutionalnetworks}.


Traditional CNNs are robust to translations (such that a translation in the input produces an equivalent translation in the feature map) but not to other transformations, such as rotations or reflections. This has motivated the generalization of CNNs to handle these symmetries \cite{cohen2016groupequivariantconvolutionalnetworks}. Data augmentation can increase robustness to these transformations, as well as prevent overfitting, as seen implemented in ImageNet \cite{10.1145/3065386}, where the training set was augmented using  horizontal reflections, amongst other transformations. However, this can be limiting in fields such as material science, where the underlying problem may obey some non-trivial transformation symmetry, which may not be learned with acceptable accuracy by simply augmenting the training set with transformed copies thereof; additionally, the lack of data availability may make augmentation of the training set infeasible. But even perfect learning of the symmetry on the training set does not guarantee generalization to the symmetry on new data, as noted by \cite{gerken2024emergentequivariancedeepensembles, dieleman2016exploitingcyclicsymmetryconvolutional, LAFARGE2021101849} and demonstrated in practice by \cite{gerken2022}. Another example where invariance to transformations (in particular, to rotations) is extremely desirable is in medical image analysis, such as in histopathology, where arbitrary orientations are common; thus, reliable models must produce consistent outputs under these (special Euclidean group $SE(2)$) transformations \cite{LAFARGE2021101849}.


To address this, the machine learning field has become interested with Equivariant Learning; from early works \cite{shawe1989building, Freeman}, to applications in feedforward networks \cite{ravanbakhsh2017equivarianceparametersharing, finzi2021practicalmethodconstructingequivariant, Ma2023UnitaryEquivariantNN} and Convolutional/Steerable networks \cite{cohen2016steerablecnns, cohen2018sphericalcnns,
bekkers2018rototranslationcovariantconvolutionalnetworks, cohen2019gaugeequivariantconvolutionalnetworks, cohen2020generaltheoryequivariantcnns, finzi2020generalizingconvolutionalneuralnetworks}. Roughly speaking, these methods constrain and build group symmetries into the model architecture and ensure that any transformation which is applied to the input
results in a corresponding and predictable transformation of the output. Indeed, suppose $G$ is a group, $\Omega \subseteq \mathbb{R}^N$ and $M, N \geq 1$ are integers. Formally, a network (or generally, a function) $\nu : \Omega \to \mathbb{R}^M$ is said to be $(T_{in},T_{out})$-equivariant (or, simply, equivariant) if and only if
\begin{equation}\label{Equivariance Relation}
    \nu(T_{in} x) = T_{out}\nu(x)\;\forall g\in G,
\end{equation}
%
and all $x\in \Omega$, where $(T_{in},T_{out})$ are two (possibly distinct) group actions of the group $G$ on $\mathbb{R}^N$ and $\mathbb{R}^M$, respectively.

Early attempts to incorporate symmetry considered constraining not the model architecture, but the network's function space. Shawe-Taylor \cite{shawe1989building} demonstrated that invariance to finite groups could be achieved by averaging outputs over the group orbit. Whilst this is sufficient for invariance, this specific approach destroys the spatial structure required for equivariant feature learning. Bao and Song \cite{bao2019equivariant} proposed ``equivarification", a method that transforms an arbitrary feedforward network into a $G$-equivariant architecture without hard-coding it in the model itself. Whilst this works without hard-coding symmetries, their method struggles to scale to larger groups. Another specialized architectural design was proposed by Ma and Chan \cite{Ma2023UnitaryEquivariantNN}, who built a feedforward network equivariant to actions of the full unitary group to predict atomic motions. However, their approach was limited by requiring the input and output group actions to be identical.

\begin{figure}[h] 
\centering
\includegraphics[width=1.0\columnwidth]{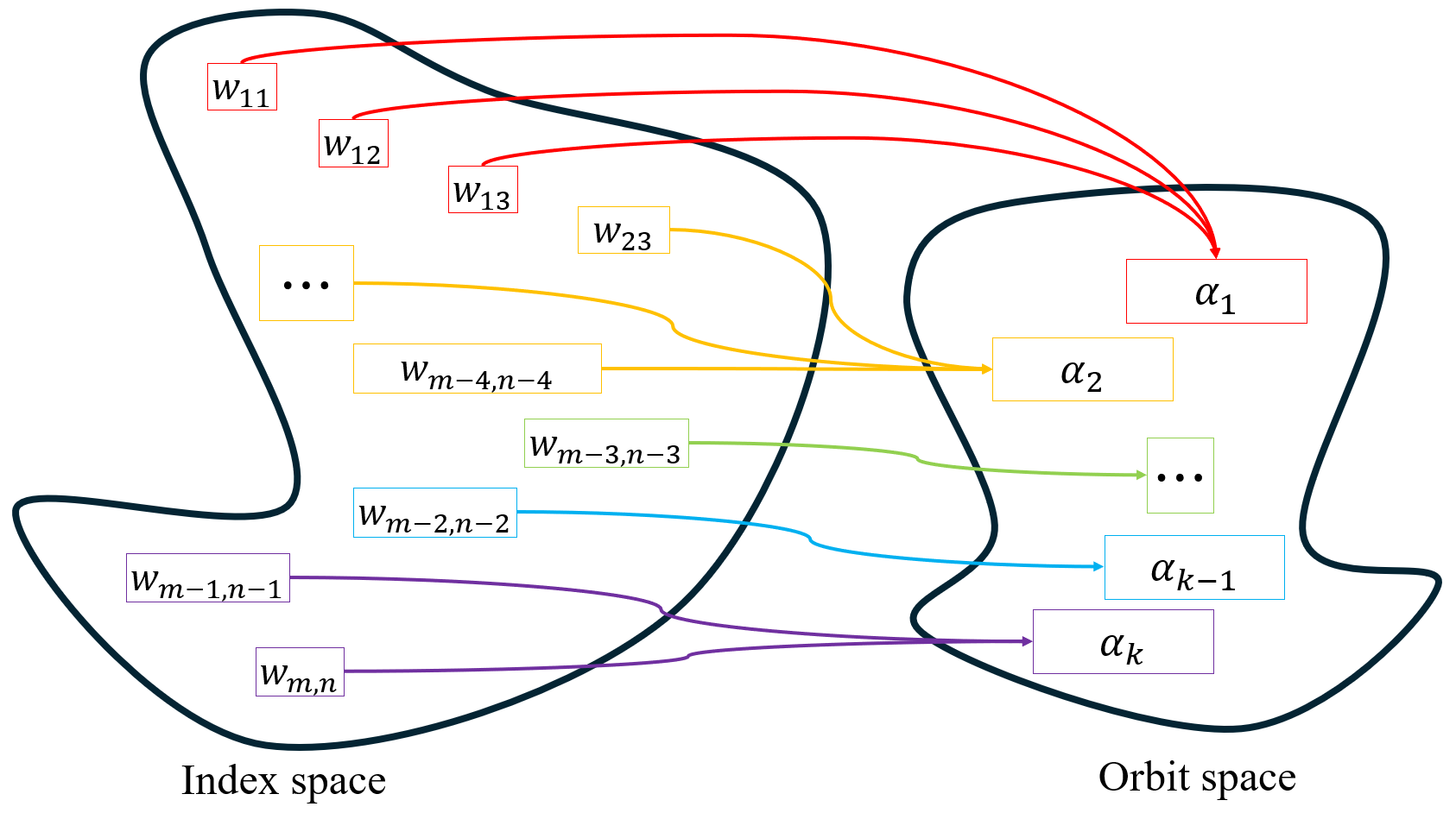}
\caption{\scriptsize{Diagram to help motivate per-orbit parameter application. Consider an equivariant weight matrix in this example, where $W\in\mathbb{R}^{m\times n}$ satisfying some constraint $\rho_{out}W = W\rho_{in}$. Indices in the same orbit share the same free parameter. We have instead of $mn$ parameters only $k<mn$.}}
\label{fig: Per Orbit Parameters}
\end{figure}

A more general framework for the construction of ENNs is based on parameter sharing resulting from group actions. Here, the group acts by transforming the input and output coordinates of the trainable parameters (such as network weights), and equivariance is enforced by constraining these parameters to intertwine appropriately with the group representations. For cyclic and permutation groups, which can be represented as permutation matrices, Ravanbakhsh et al. \cite{ravanbakhsh2017equivarianceparametersharing} noted that the weight matrix $W$ in a neural network layer $\phi_l:\mathbb{R}^I \to \mathbb{R}^O$, where $\phi_l (z_l) := \sigma_l(W_l z_{l-1})$ with $\sigma_l$ denoting an activation function, $z_{l-1}\in\mathbb{R}^I$ is the input from the previous layer $l-1$, and $W_l\in\mathbb{R}^{\mathfrak{O}\times I}$, is ($\rho_{in},\,\rho_{out}$)-equivariant if
\begin{equation*}
    \rho_{out}(g)W_l = W_l\rho_{in}(g),
\end{equation*}
where $\rho_{in}(g)\in\mathbb{R}^{I\times I}$, $\rho_{out}(g)\in\mathbb{R}^{\mathfrak{O}\times \mathfrak{O}}$ are group representations of $G$. Note that if the activation function $\sigma_l$ is a pointwise function (applied to each component of the vector in $\mathbb{R}^I$ individually), then $\mathbb{R}^{\mathfrak{O}} = \mathbb{R}^O$. The matrix $W_l$ is also sometimes referred to as an intertwiner of the two representations \cite{serre1977representations}.
%
%
%
%
%
As the above identity must hold for all elements $g$ in the underlying group $G$, elements of the weight matrix become tied with one another, thereby reducing its size complexity (as visualized in Figure (\ref{fig: Per Orbit Parameters})) and potentially training time. Moreover, these weight ties hard-code equivariance into the neural network layer itself.

Finzi et al. \cite{finzi2021practicalmethodconstructingequivariant} introduce a general framework for building networks equivariant to any matrix group, such as the Lorentz group $O(1,3)$ and the Rubik's cube group, by providing a general algorithm for solving for equivariant layers. They show that the equivariance constraints on the weights can be reduced to a finite set of $M + D$ linear constraints (where $M$ is the number of discrete generators and $D$ is the dimension of the group). To solve for the equivariant layers, their method uses singular value decomposition (SVD) to solve for the null space of $\rho_{out}(g) \otimes \rho_{in}(g^{-1})^T - \mathbb{I}$ derived from the weight-tie constraint $\rho_{out}(g) W  = W\rho_{in}(g)$:
\begin{align*}
     \rho_{out}(g) W \rho_{in}(g)^{-1} =&\;W\;\forall g \in G\\
     \rho_{out}(g) \otimes \rho_{in}(g^{-1})^T \text{vec}(W) =&\;\text{vec}(W)\\
     (\rho_{out}(g) \otimes \rho_{in}(g^{-1})^T- \mathbb{I}) \text{vec}(W)& = 0,
\end{align*}
where $\text{vec}(W)$ denotes vectorisation of the weight matrix $W$. They demonstrate that this method recovers well-known architectures, such as vanilla CNNs (translation equivariance: $G = \mathbb{Z}_n \times \mathbb{Z}_n$), Deep Sets \cite{zaheer2018deepsets} (permutation equivariance: $G=S_n$), and GCNNs \cite{cohen2016groupequivariantconvolutionalnetworks} (translation with $\frac{\pi}{2}$-rotation equivariance: $G = \mathbb{Z}_4 \ltimes (\mathbb{Z}_n \times \mathbb{Z}_n)$). They also further highlight that linear layers and pointwise activations such as ReLU do not give universal networks for certain groups. To achieve universality, they incorporate a bilinear layer, using tensor products (multiplication of features).

Beyond the applications of equivariance to NN-based architectures, several other works focus on CNN-based models. An early work, Group Equivariant CNNs (G-CNNs) \cite{cohen2016groupequivariantconvolutionalnetworks}, handles discrete symmetries to groups p4 (collection of translations and rotations by $\frac{\pi}{2}$ about any centre of rotation on a square grid) and p4m (p4 with the addition of mirror reflections). Instead of convolving filters purely over a pixel grid, G-CNNs utilize ``G-convolutions" that convolve feature maps over a specific symmetry group. By replicating and rotating filters, G-CNNs achieve exact equivariance and greater weight sharing (as they are re-used across different orientations) without increasing the number of trainable parameters. Applications to continuous symmetries ($SO(2)$ and $SO(3)$) have been explored by  \cite{cohen2016steerablecnns, weiler20183dsteerablecnnslearning}. Rather than replicating filters, S-CNNs restrict their convolution kernels to a continuous, steerable mathematical basis, such as spherical harmonics. This allows rotations to be applied analytically. However, because the infinite basis must be truncated in practice, S-CNNs typically yield only approximate equivariance (which can be sufficient enough for many tasks).

Although the literature on equivariant models spans diverse neural architectures, the present work concentrates exclusively on equivariant feedforward networks with respect to cyclic groups.

\subsubsection{Weight Sharing}

In standard (non-equivariant) networks, every feature or node within a layer is treated independently and no relationships between them assumed. But in equivariant neural networks, this is different and several nodes can be related to one another. The nodes rely on orbits, which dictate how parameters are shared. Suppose a symmetry group $G$ acts on the feature space of a network through a matrix representation. For any given feature index $i$, its orbit is defined as the set of all indices it can be mapped to by the actions of the group 
\begin{equation*}
    \mathcal{O}_i = \{g^p i : g \in G\; \text{and}\; p = 1, \hdots, P\},
\end{equation*}
where each group element $g\in G$ corresponds to one specific matrix under this representation with juxtaposition denoting the group action of $G$ on the index set. At $p=P$, the original index is recovered:  $g^P i = i$. One can understand this process as the group actions partitioning the feature space into disjoint equivalence classes which are the orbits. For the network to be equivariant, feature nodes belonging to the same orbit are essentially equivalent and must be processed identically. In a similar way, the weight matrix of some given layer is constrained by two representations: the input and output representations that appear in the intertwiner identity $\rho_{out}W = W\rho_{in}$. The vector containing the features $z$ on the other hand only depend on one representation, $\rho_{out}$. 

Generally, the space of orbits is smaller than the index space, unless $\rho_{in},\rho_{out}$ are chosen to be identity matrices, which would result in an unconstrained $W$. For $W$ to  satisfy equivariance, indices within a unique orbit are assigned one free parameter. This demonstrates the potential of hard-coding equivariance into the model architecture as a strategy for reducing the number of trainable parameters.

For building layers equivariant to an arbitrary group, one can solve for the null space of the constraint matrix as outlined in \cite{finzi2021practicalmethodconstructingequivariant}. However, in the case where the representations are permutation matrices, the task of constraining the weights is more straightforward, and can be implemented by tying parameters across orbits \cite{ravanbakhsh2017equivarianceparametersharing}.
%
%



{ Let $W\in\mathbb{R}^{O\times I}$ be a weight matrix of a network layer. Suppose that possibly two different group actions of elements of the cyclic group of order $k$, $g\in G=C_k$ (sometimes referred to by $\mathbb{Z}_k$), act on the input space $\mathbb{R}^{I\times I}$ and output space $\mathbb{R}^{O \times O}$ through the permutation matrix representations
\begin{align*}
    \rho_{in}: \;& G \to GL(I)\\
    \rho_{out}: \;& G \to GL(O).
\end{align*}
The weight matrix is constrained by 
\begin{equation}\label{Eq: W constraint}
    \rho_{out}(g) W = W \rho_{in}(g)\;\forall g \in G,
\end{equation}
which can be rewritten as a constraint on the indices of $W$. If $g\in C_k$ is the generator element of the group, then the constraint (\ref{Eq: W constraint}) holds for all powers of $g$ and hence for all of $C_k$. Let $\sigma: \{1, \hdots, I\}\to  \{1, \hdots, I\}$ be the permutation corresponding to $\rho_{in}$ and $\pi: \{1, \hdots, O\}\to  \{1, \hdots, O\}$ be the permutation corresponding to $\rho_{out}$, with 
\begin{align*}
    \rho_{in} v &= u : u_i = v_{\sigma(i)}\\
    \rho_{out} v &= u : u_i = v_{\pi(i)}.
\end{align*}
Then the LHS of (\ref{Eq: W constraint}) can be written as
\begin{equation*}
    (\rho_{out}(g) W)_{ij} = \sum_k (\rho_{out})_{ik} W_{kj} = W_{\pi(i), j},
\end{equation*}
and the RHS of  (\ref{Eq: W constraint}) as
\begin{equation*}
    (W\rho_{in}(g))_{ij} = \sum_k  W_{ik} (\rho_{in})_{kj} = W_{i, \sigma^{-1}(j)}.
\end{equation*}
By re-indexing $j \to \sigma(j)$, we then have
\begin{equation*}\label{Eq: W constraint on indices}
    W_{\pi(i), \sigma(j)} = W_{ij}
\end{equation*}
$\forall i \in \{1, \hdots, O\}$ and $\forall j \in \{1, \hdots, I\}$. Orbits of this action partition the index set of $W$ ($(i,j)\in\{1, \hdots, O\}\times \{1, \hdots, I\}$):
\begin{equation*}
    \mathcal{O}(i,j) = \{(\pi^p(i), \sigma^p(j)): p = 1, 2, 3, \hdots, P\},
\end{equation*}
where $p=P$ permutes the indices back to their original state $(\pi^P(i), \sigma^P(j)) = (i,j)$.} Therefore, $\mathcal{O}(i,j)$ is the collection of all indices of $W$ that are linked to the pair $(i,j)$ across the permutations $\pi$ and $\sigma$.

All entries belonging to the same orbit must take the same value in order to satisfy the equivariance constraint (\ref{Eq: W constraint}), allowing one to assign a single free parameter to each orbit. Then, the resulting layer map containing the matrix $W$ is equivariant with respect to the cyclic group actions. Most importantly, the number of independent parameters is reduced from $mn$ to the number of disjoint orbits, which is typically is much smaller.



\section{Methodology}

\subsection{Network Architecture}

In our equivariant network architectures, the weight matrices are constrained by representation matrices of cyclic groups of different orders. We divide these into two categories: ones we construct that act as approximate rotations (via permutation of the pixels of an image), and those that are shift matrices (regular representations). We distinguish the constructed permutation matrices as $\varrho_{C_k}(g)$ and the regular representations as $\rho_{C_k}(g)$ of a given cyclic group $C_k$. {We will later also use an additional subscript for both representation types to denote the layer they are applied in.}  The method of construction of the $\varrho_{C_k}(g)$-matrices is described in Appendix \ref{Construction of the Rotation-like Permutation Matrix}.

\begin{figure*}[t] 
\centering
\includegraphics[width=\textwidth]{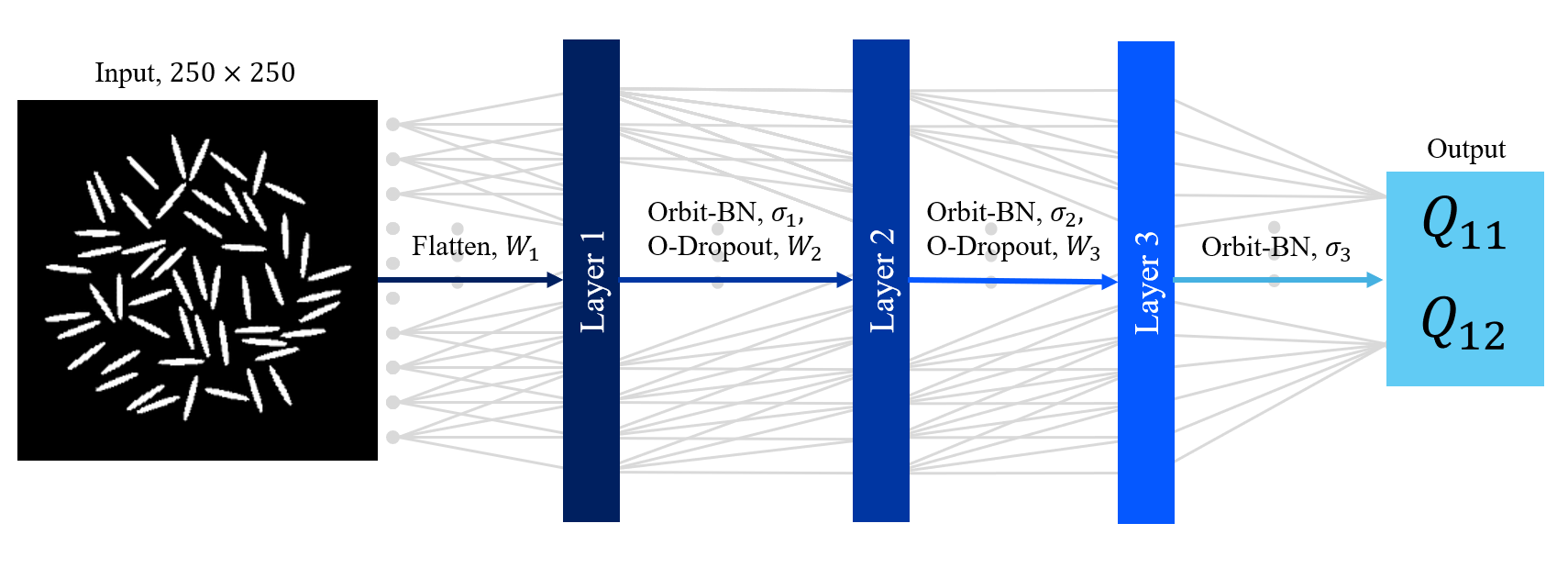}
\caption{{Model architecture of our equivariant networks. The input images are vectorized row-wise and are strictly black and white (represented by 0s and 1s). Through a sequence of layers the image is mapped to a 2D vector, which corresponds to the associated $Q$-tensor values $Q_{11}, Q_{12}$. The full 2D $Q$-tensor can be derived from these two independent variables. The non-equivariant architectures are equivalent in design, and use regular Batch norm (BN) and Dropout. }}
\label{fig: Model architecture}
\end{figure*}

All neural networks in the work have the same 3-layer form (see Figure \ref{fig: Model architecture})
\begin{equation}\label{Form of ENN}
    \nu(x) = \sigma_3 \Bigl( W_3 \, \sigma_2 \Bigl( W_2 \, \sigma_1 \bigl( W_1 x \bigr) \Bigr) \Bigr) \in\mathbb{R}^2,
\end{equation}
with $x\in\mathbb{R}^{250^2}$, where after each application of the weight matrices, an orbit-based batch normalization step is performed, and after the application of $\sigma_1$ and $\sigma_2$, orbit-based dropout occurs (find further details on equivariant regularization in \ref{Equivariant Regularization}). In the final layer, $\sigma_3$ maps to the output via a non-learnable matrix $\mathfrak{L}$ (\ref{General Intertwiner L}). Note that the biases in (\ref{Form of ENN}) are not present: this is because orbit-based batch norm is applied, which effectively cancels these out. However, orbit-based batch normalisation introduces, similarly to standard batch norm, scaling and shifting parameters ($\gamma$ and $\beta$), of which the shifting values act as the biases.

The activation functions used are 
\begin{align*}
    \sigma_1(z) = \sigma_2(z) &:=  GELU(x) = z\Phi(z)\\
    \sigma_3(z) &:= \mathfrak{L}z \frac{\tanh(||\mathfrak{L}z||_2)}{2||\mathfrak{L}z||_2 },
\end{align*}
where $\tanh$ is the hyperbolic tangent function, $||\cdot||_2$ denotes the Euclidean norm, and $GELU$ is the Gaussian Error Linear Unit activation \cite{hendrycks2023GELU}, where $\Phi(z)$ is the Cumulative Distribution Function of the standard normal distribution:
\begin{align*}
    \Phi(z) &= \frac{1}{\sqrt{2\pi}}\int_{-\infty}^z e^{-t^2/2}dt\\
    &\approx \frac{1}{2} (1 + \tanh(\sqrt{\frac{2}{\pi}}(z+0.044715z^3))).
\end{align*}
Because all representations used are permutation matrices, any pointwise activation function is equivariant and satisfies the condition for $\sigma_1, \sigma_2$ (\ref{Eq: Activation constraints}) required for equivariance. Although, the activation $\sigma_3(z)$ is a scaling function, and therefore linear, the overall network remains nonlinear since the $GELU$-activations are themselves nonlinear.


\subsubsection{Equivariant Architectures}

\begin{figure}[h] 
\centering
\includegraphics[width=1.0\columnwidth]{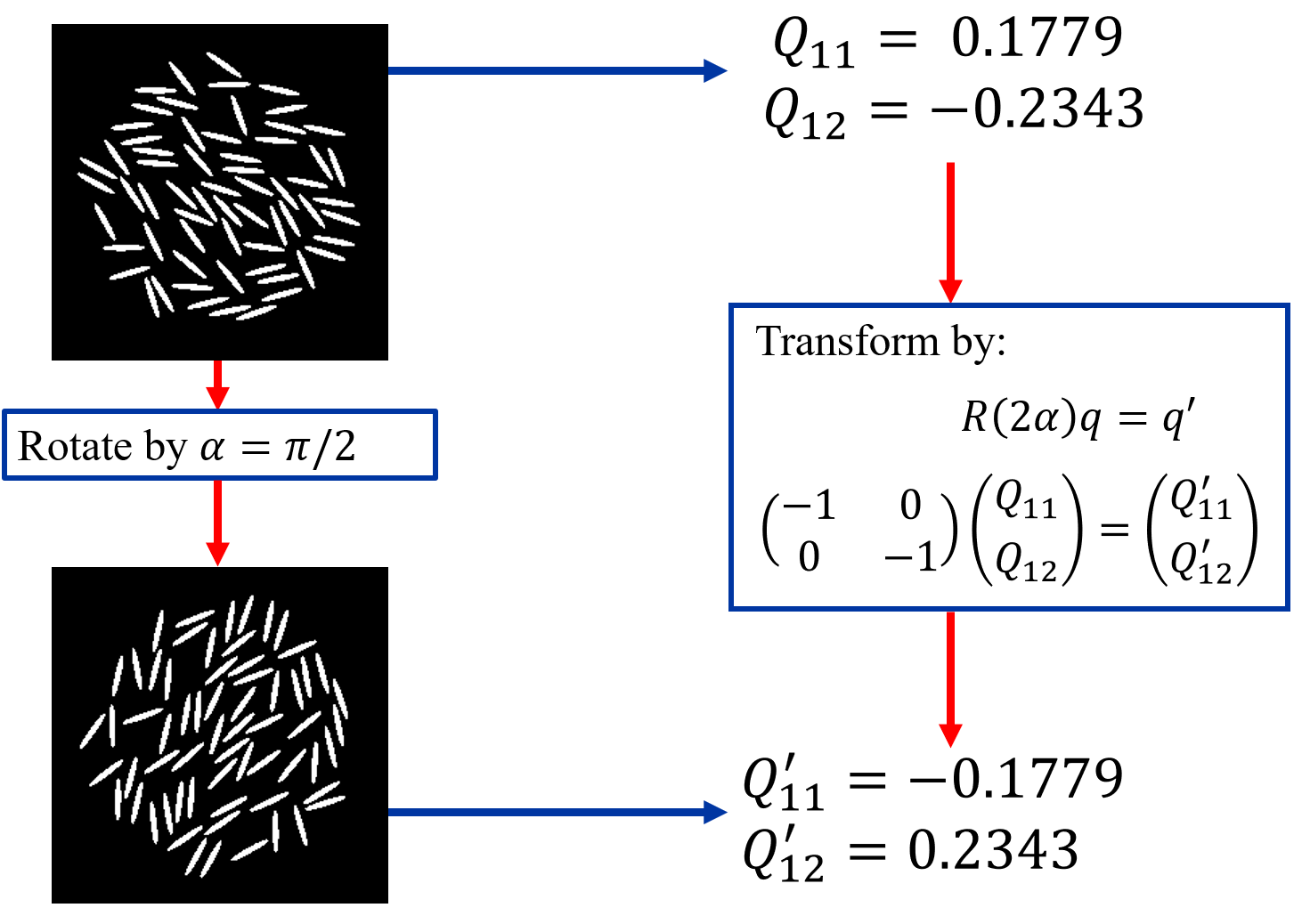}
\caption{\scriptsize{Diagram of the needed equivariance constraint for $\alpha = \frac{\pi}{2}$ (when $k=4$). Generally, we require that a $\theta = \frac{2\pi}{k}$ rotation-like transformation of the (vectorized) input, by $\varrho_{C_k,\;0}(g)$, corresponds to the output vector transforming under the two-dimensional rotation matrix, evaluated at $2\theta =\frac{4 \pi}{k}$.}}
\label{fig: Equivariance_of_Q}
\end{figure}

The neural networks (\ref{Form of ENN}) will obey the $C_k$-equivariance constraint below by construction and predict $\nu \approx \vec{q}$ as defined in (\ref{Eq: Vector form of Q equivariance}). Specifically, the network must satisfy
\begin{equation}\label{Equivariant NN}
    \nu(\varrho_{C_k,\;0}(g^p) x) = R_{\frac{4p\pi}{k}} \nu(x),
\end{equation}
where $g$ is the generator element of $C_k$, and with $p = 1, \hdots k$. This is only true if the weights satisfy the conditions (\ref{W1 Intertwiner}) and (\ref{W2, W3 Intertwiner}). The term $\varrho_{C_k,\;0}(\cdot)$ acts as the input representation (in layer 0) and $R_{\frac{4p\pi}{k}}$ as the output representation. Figure \ref{fig: Equivariance_of_Q} demonstrates this in the case for the cyclic group $C_4$ (rotations of $\frac{\pi}{2}$) with $\varrho_{C_4,\;0}(g)$ and $R_{\frac{4\pi}{4}} = R_{\pi}$ as the input/output representations of $G=C_k$.
To demonstrate this equivariance constraint on the network $\nu(x)$ holds, we show how the LHS of (\ref{Equivariant NN}) can be expressed as its RHS, and how conditions on the weight-ties and equivariant activations arise. Since all networks share the same 3-layer architecture (\ref{Form of ENN}) we have the following equivariance chain:
\begin{align}\label{Full equivariance chain}
    \begin{split}
    \nu(\varrho_{C_k,\;0}(g^p) x) &= \sigma_3\left(W_3 \sigma_2\left(W_2 \sigma_1\left(W_1 \varrho_{C_k,\;0}(g^p) x\right)\right)\right)\\[6pt]
    &= \sigma_3\left(W_3 \sigma_2\left(W_2 \sigma_1\left(\varrho_{C_k,\;1}(g^p)W_1  x\right)\right)\right)\\[6pt]
    &= \sigma_3\left(W_3 \sigma_2\left(W_2 \varrho_{C_k,\;1}(g^p)\sigma_1\left(W_1  x\right)\right)\right)\\[6pt]
    &= \sigma_3\left(W_3 \sigma_2\left(\rho_{C_k,\;2}(g^p)W_2 \sigma_1\left(W_1  x\right)\right)\right)\\[6pt]
    &= \sigma_3\left(W_3 \rho_{C_k,\;2}(g^p) \sigma_2\left(W_2 \sigma_1\left(W_1  x\right)\right)\right)\\[6pt]
    &= \sigma_3\left(\rho_{C_k,\;3}(g^p)W_3  \sigma_2\left(W_2 \sigma_1\left(W_1  x\right)\right)\right)\\[6pt]
    &= R_{\frac{4 p \pi}{k}}\sigma_3\left(W_3  \sigma_2\left(W_2 \sigma_1\left(W_1  x\right)\right)\right)\\[6pt]
    &= R_{\frac{4p\pi}{k}} \nu(x).
    \end{split}
\end{align}
Each pair of consecutive rows in (\ref{Full equivariance chain}) must be equal for equivariance to hold. This therefore requires the following conditions on both the weights and activations:
\begin{align}
\begin{split}\label{Eq: Weight Constraints}
    \varrho_{C_k,\;1}(g^p)W_1 &=  W_1 \varrho_{C_k,\;0}(g^p)  \\[6pt]
    \rho_{C_k, 2}(g^p)  W_2 &= W_2  \varrho_{C_k,\;1}(g^p)  \\[6pt]
    \rho_{C_k, 3}(g^p)  W_3 &= W_3 \rho_{C_k, 2}(g^p)
\end{split}
\end{align}
and
\begin{align}\label{Eq: Activation constraints}
    \begin{split}
       \sigma_1(\varrho_{C_k,\;1}(g^p) x)  &= \varrho_{C_k,\;1}(g^p) \sigma_1(x) \\[6pt]
       \sigma_2(\rho_{C_k, 2}(g^p) y)  &= \rho_{C_k, 2}(g^p) \sigma_2(y) \\[6pt]
       \sigma_3(\rho_{C_k, 3}(g^p)  z)  &= R_{\frac{4p\pi}{k}} \sigma_3(z), 
    \end{split}
\end{align}
for some feature vectors $x,\,y$, and $z$. The equivariance chain holds for all $g^p$ since it holds for the generator element $g\in C_k$, as mentioned above.

\subsubsection{Propagation of Equivariance Through Weight-Constrained Layers}



In these $C_k$-equivariant networks, the group action at each layer is encoded as a representation that constrains the learnable weights. By Schur's lemma \cite{serre1977representations}, equivariant linear maps can only connect matching irreducible representations (irreps) between adjacent layers. Information associated with a given irrep can only flow forward if that irrep is present in every layer's representation along the path to the output.

Note that the final (output) representation, $R_{\frac{4p\pi}{k}}$, can be further decomposed into a conjugate pair of complex irreps.  If any intermediate layer's representation lacks either of these irreps, the equivariant weight constraint forces the corresponding channel to zero and no learnable parameterization ($W$ mapping) can recover it downstream.

The requirement on intermediate representations is therefore straightforward: each must be a valid $C_k$-representation whose irrep decomposition contains at least one copy of each irrep required by the output. The regular representation (cyclic permutation of a $k$-dimensional vector by some $\mathbb{R}^{k\times k}$ representation) is always a safe default, as it contains every irrep of $C_k$ exactly once. Our custom rotation-like ($\varrho$) representations, which may be larger than the standard regular representation matrices: $\mathbb{R}^{K\times K}: K>k$, also satisfy this even though they may contain multiple copies of each irrep.


Unlike the $\varrho$-type matrices, the regular representations decompose as a direct sum ($\oplus$) of the roots of unity satisfying $z^k=1$ ($m_1 \chi_1 \oplus \hdots \oplus m_k \chi_k$) with multiplicities $m_1, \hdots, m_k =1$ exactly, where $\chi_i$ represent the irreps.

We discuss in the next section what form these representations take in each $C_k$-equivariant model.

\subsection{Equivariant Representations}

\subsubsection{Rotation-like representations}

The first type is where the cyclic representations, corresponding to $C_k$, act as $\frac{2\pi}{k}$ rotations. This means that they permute the input (vectorized square image) in such a way that when the output is reconstructed back into an image, it appears as an approximately rotated version of the input. Note that when applying these transformations, the image must be defined on a circular domain of the square image, except for the $C_4$ case, where the image can be defined on a square grid. This is because the application of these matrices ``fixes" the corners so that the transformations appears as a ``rotation" of the circular subspace. 


We define the representations for the cyclic groups 
$$C_k = \langle g: g^k = e\rangle$$
as $\varrho_{C_k}(g): C_k \to GL(V)$, where $V$ is a vector space of dimension equal to the vectorized input image, with $\varrho_{C_k}(g^p) = \varrho_{C_k}(g)^p$ where when $p=k$, $\varrho_{C_k}(g)^k = \mathbb{I}$ (identity matrix in $GL(V)$). The weight matrices connecting the input layer (layer 0) to layer 1 (via $W_1$) of all networks are constrained by the following representations:
\begin{equation}\label{W1 Intertwiner}
     \varrho_{C_k,\;1}(g^p)W_1 = W_1 \varrho_{C_k,\;0}(g^p)
\end{equation}
for $p = 1, \hdots, k$. Table \ref{tab:W1-constraints} displays these representations for each $C_k$-equivariant architecture.
\begin{table}[h]
\centering
\caption{\scriptsize{The table presents all representations constraining $W_1$, satisfying (\ref{W1 Intertwiner}), that are used in each $C_k$-model. Note that $\varrho_{C_{128},\;1}(g)$ and $\varrho_{C_{256},\;1}(g)$ map to larger spaces; these are the smallest sizes for which the algorithm could construct a $\frac{2\pi}{128}$ and $\frac{2\pi}{256}$ matrix representation, respectively (correspond to the rotations of $14\times 14$ and $20\times 20$ pixel images). }}
\label{tab:W1-constraints}
\begin{ruledtabular}
\scriptsize{
\begin{tabular}{ccc}
$C_k$-Model & $\varrho_{C_k,\;0}(g)$ & $\varrho_{C_k,\;1}(g)$ \\
\hline
$C_{4}$ & $\varrho_{C_{4},\;0}(g):\;C_{4} \to GL(250^2)$ &   $\varrho_{C_{4},\;1}(g):\;C_{4} \to GL(100)$  \\
$C_{8}$ & $\varrho_{C_{8},\;0}(g):\;C_{8} \to GL(250^2)$ &  $\varrho_{C_{8},\;1}(g):\;C_{8} \to GL(100)$ \\
$C_{16}$ & $\varrho_{C_{16},\;0}(g):\;C_{16} \to GL(250^2)$ & $\varrho_{C_{16},\;1}(g):\;C_{16} \to GL(100)$ \\
$C_{32}$ & $\varrho_{C_{32},\;0}(g):\;C_{32} \to GL(250^2)$ & $\varrho_{C_{32},\;1}(g):\;C_{32} \to GL(100)$ \\
$C_{64}$ & $\varrho_{C_{64},\;0}(g):\;C_{64} \to GL(250^2)$ & $\varrho_{C_{64},\;1}(g):\;C_{64} \to GL(100)$ \\
$C_{128}$ & $\varrho_{C_{128},\;0}(g):\;C_{128} \to GL(250^2)$ & $\varrho_{C_{128},\;1}(g):\;C_{128} \to GL(196)$ \\
$C_{256}$ & $\varrho_{C_{256},\;0}(g):\;C_{256} \to GL(250^2)$ & $\varrho_{C_{256},\;1}(g):\;C_{256} \to GL(400)$ 
\end{tabular}}
\end{ruledtabular}
\end{table}

\subsubsection{Regular representations}

The weight matrices connecting layer 1 to layer 2 (by $W_2$) and those that connect layer 2 to the (final hidden) layer 3 (using $W_3$) are constrained by the following permutation matrices:
\begin{align}\label{W2, W3 Intertwiner}
\begin{split}
      \rho_{C_k, 2}(g^p)  W_2 &= W_2  \varrho_{C_k,\;1}(g^p)  \\
  \rho_{C_k, 3}(g^p)  W_3 &= W_3 \rho_{C_k, 2}(g^p),
\end{split}
\end{align}
where $ \rho_{C_k, 2}(g^p), \;\rho_{ C_k, 3}(g^p)$ are implemented as cyclic shift matrices (regular representations of $C_k$). Note that $ \rho_{ C_4, 3}(g^p)$ is a special case; it is the direct sum of two smaller $4\times 4$ such (cyclic) matrices.


\begin{table}[h]
\centering
\caption{\scriptsize{The table presents the (regular) representations constraining the weights $W_2$ and $W_3$, satisfying (\ref{W1 Intertwiner}), used in each $C_k$-model.}}
\label{tab:W2,W3-constraints without L}
\begin{ruledtabular}
\scriptsize{
\begin{tabular}{ccc}
$C_k$-Model & $\rho_{C_k,\;2}(g)$ & $\rho_{C_k,\;3}(g)$  \\
\hline
$C_{4}$ & $\rho_{C_{4},\;2}(g):\;C_{4} \to GL(8)$ &   $\rho_{C_{4},\;3}(g):\;C_{4} \to GL(8)$ \\
$C_{8}$ & $\rho_{C_{8},\;2}(g):\;C_{8} \to GL(8)$ &  $\rho_{C_{8},\;3}(g):\;C_{8} \to GL(4)$ \\
$C_{16}$ & $\rho_{C_{16},\;2}(g):\;C_{16} \to GL(16)$ & $\rho_{C_{16},\;3}(g):\;C_{16} \to GL(8)$ \\
$C_{32}$ & $\rho_{C_{32},\;2}(g):\;C_{32} \to GL(32)$ & $\rho_{C_{32},\;3}(g):\;C_{32} \to GL(16)$ \\
$C_{64}$ & $\rho_{C_{64},\;2}(g):\;C_{64} \to GL(64)$ & $\rho_{C_{64},\;3}(g):\;C_{64} \to GL(32)$ \\
$C_{128}$ & $\rho_{C_{128},\;2}(g):\;C_{128} \to GL(128)$ & $\rho_{C_{128},\;3}(g):\;C_{128} \to GL(64)$ \\
$C_{256}$ & $\rho_{C_{256},\;2}(g):\;C_{256} \to GL(256)$ & $\rho_{C_{256},\;3}(g):\;C_{256} \to GL(128)$ 
\end{tabular}
}
\end{ruledtabular}
\end{table}

\subsubsection{The Intertwiner $\mathfrak{L}$}

The final layer of our $C_k$-equivariant networks must be at least of size $\frac{k}{2}$.

We can give a general form of the intertwiner $\mathfrak{L}\in\mathbb{R}^{2\times k/2}$, which is required to satisfy:
\begin{equation}\label{L-Intertwiner}
\begin{split}
    \mathfrak{L}\rho_{C_k,\,3}(g^p) &= R_{2\frac{2p\pi}{k}}\mathfrak{L} \\
                                  &= R_{\frac{4p\pi}{k}}\mathfrak{L},
\end{split}
\end{equation}
for $p=1, \hdots, k $, where $R_{\frac{4p\pi}{k}}$ comes from the equivariance condition of the $C_k$-equivariant networks,
\begin{equation*}
    \nu(\varrho_{C_k,\;0}(g^p) x) = R_{\frac{4p\pi}{k}} \nu(x).
\end{equation*}
%

We define these intertwining $\mathfrak{L}$-matrices as
\begin{equation}\label{General Intertwiner L}
    \mathfrak{L} = [l_{ij}] :
\begin{cases}
    l_{1,j} = \cos\left(\frac{4\pi j}{k}\right) \\
    l_{2,j} = \sin\left(\frac{4\pi j}{k}\right)
\end{cases}
\;\; \text{for } j = 0, \dots, \frac{k}{2}-1,
\end{equation}
for even $k$. Each group $C_k$ determines the rotation angle $\theta =  \frac{2p\pi}{k}$ (represented by $\rho_{C_k,\,3}(g^p)$) which in return gives rise to the associated output rotation matrix $R_{2\theta} = R_{\frac{4p\pi}{k}}$.

\begin{table}[h]
\centering
\caption{{\scriptsize The non-learnable intertwining matrix $\mathfrak{L}$ that maps from the final layer to the output in $\mathbb{R}^2$. It is part of the final activation function $\sigma_3.$}}
\label{tab: L}
\begin{minipage}{0.6\columnwidth} 
\begin{ruledtabular}
{
\begin{tabular}{cc}
$C_k$-Model & $\mathfrak{L}$\\
\hline
 $C_{4}$ &   $\mathfrak{L}:\;\mathbb{R}^{8} \to \mathbb{R}^2$\\
 $C_{8}$ &  $\mathfrak{L}:\;\mathbb{R}^{4} \to \mathbb{R}^2$\\
 $C_{16}$ &   $\mathfrak{L}:\;\mathbb{R}^{8} \to \mathbb{R}^2$\\
 $C_{32}$ &   $\mathfrak{L}:\;\mathbb{R}^{16} \to \mathbb{R}^2$\\
 $C_{64}$ &   $\mathfrak{L}:\;\mathbb{R}^{32} \to \mathbb{R}^2$\\
 $C_{128}$ &   $\mathfrak{L}:\;\mathbb{R}^{64} \to \mathbb{R}^2$\\
 $C_{256}$ &   $\mathfrak{L}:\;\mathbb{R}^{128} \to \mathbb{R}^2$\\
\end{tabular}
}
\end{ruledtabular}
\end{minipage}
\end{table}

\subsection{Equivariant Activations and Regularization Techniques}

\subsubsection{Equivariant Activations}

In general, commonly-used activations, such as ReLU or tanh, break equivariance so that for two arbitrary group representations, $\pi_1$ and $ \pi_2$, they are such that $\sigma(\pi_1(g) y )\neq \pi_2(g) \sigma(y)$ for some vector $y$. For example, consider the Rectified Linear Unit (ReLU) activation
\begin{align*}
    \sigma(y) &= \text{ReLU}(y)\\
    &=(\text{max}(0, y_0), \text{max}(0, y_1), \hdots, \text{max}(0, y_i))
\end{align*}
and define the two {geometric} representations be the same $\frac{\pi}{2}$ rotation matrix
\begin{equation*}
    \pi_1(g) = \pi_2(g) = 
    {\begin{pmatrix}
    -1 & 0 \\
    0 & -1
     \end{pmatrix}}.
\end{equation*}
One can confirm that $\sigma(\pi_1(g) y ) \neq \pi_2(g)\sigma( y )$ in this example, so equivariance generally does not hold. We can however achieve $\frac{\pi}{2}$-equivariance by using permutation representations. Then element-wise activations, such as ReLU or tanh, preserve equivariance. This is because permutation representations change a value's location, but not its value, unlike with the rotation matrix above. Thus, in our networks, we use the GELU activation for $\sigma_1$ and $\sigma_2$
\begin{equation}\label{sigma1, sigma2}
    \sigma_1(y) =  \sigma_2(y) = \text{GELU}(z).
\end{equation}
In the final layer, we construct an activation function that preserves the equivariance requirement: 
\begin{equation*}
    \sigma_3(\rho_{C_k,\;3}(g^p)z) = R_{\frac{4p\pi}{k}}\sigma_3(z),
\end{equation*}
and define the map to the output as
\begin{equation}\label{sigma3}
    \sigma_3(z) : = \frac{\tanh(||\mathfrak{L}z||_2)}{2||\mathfrak{L}z||_2} \mathfrak{L}z .
\end{equation}
%
The activation is a bounded function that maps any input $z$ to a $\mathbb{R}^2$ vector where each component of $\sigma_3$ satisfies $-0.5< \sigma_3(z)_i<0.5$ ($i=1,2$). This activation is equivariant, viz.
\begin{align*}
    \sigma_3(\rho_{C_k,\;3}(g^p)z) &= \mathfrak{L}\rho_{C_k,\;3}(g^p)z \frac{\tanh(||\mathfrak{L}\rho_{C_k,\;3}(g^p)z||_2)}{2||\mathfrak{L}\rho_{C_k,\;3}(g^p)z||_2}\\
    &= \mathfrak{L}\rho_{C_k,\;3}(g^p)z \frac{\tanh(||\mathfrak{L}z||_2)}{2||\mathfrak{L}z||_2}\\
    &= R_{\frac{4p\pi}{k}}\mathfrak{L}z \frac{\tanh(||\mathfrak{L}z||_2)}{2||\mathfrak{L}z||_2}\\
    &= R_{\frac{4p\pi}{k}}\sigma_3(z),
\end{align*}
noting that $\mathfrak{L}$ is an intertwiner between the input and output representations as needed by (\ref{L-Intertwiner}) and constructed in (\ref{General Intertwiner L}). More on norm-based activations can be found in the thesis of Cesa \cite{cesa_e2_steerable_cnns} (see p. 49, 92).

\subsubsection{Equivariant Regularization}\label{Equivariant Regularization}

Batch normalization (BN) \cite{ioffe2015batchnormalizationacceleratingdeep} is a method used to stabilise the loss landscape and speed-up model training by normalizing the inputs prior to an activation function. Standard BN, however, does not preserve equivariance. To show this, consider a symmetry group $G$ acting on feature indices $\{1, 2, \hdots, N\}$. Regular BN breaks the symmetry by how it processes every feature index $j$ independently. Statistics $\mu_j$ and $\sigma_j$ are calculated and scale $\gamma$ and shift  $\beta$ parameters are learned per unique index. Under the group action two or more features may be related (i.e., be in the same orbit) and must therefore be treated identically, yet are treated differently by the BN algorithm. Therefore, we would find that for a group action $g\in G$ the BN map is not equivariant, i.e., BN$(g \cdot z)\neq g \cdot$BN$(z)$ for all $g\in G$.

To resolve this, we modify the BN method so that it operates on orbits, denoting this new method by OrbitBN. Instead of treating every feature index uniquely, we compute statistics and learn scale/shift parameters per orbit (we share the same parameters for indices per orbit). By forcing all features in the same orbit to undergo the same normalization transformation, the modified BN preserves the structure needed for equivariance: OrbitBN$(g \cdot z) = g \cdot$OrbitBN$(z)$.

In the networks, $z$ is a feature vector at a given layer and group elements $g\in G$ are represented as matrices. Since we apply OrbitBN after multiplication with the equivariant weights $W$, which satisfies some form of $\rho_{out} W = W \rho_{in}$, the features will always be constrained by the layer's output representations' ($\rho_{out}$'s) orbits.

For example, consider an arbitrary layer $l$ with weight matrix $W_l$ constrained by some input/output representations $\rho_{out, l}W_l=W_l\rho_{in, l}$. We require that the orbit-based BN method satisfies
\begin{align*}
    \text{OrbitBN}(\rho_{out, l}z_l) = \rho_{out, l}\text{OrbitBN}(z_l),
\end{align*}
where $z_l= W_l z_{l-1}$ is the transformed input from the previous layer. We compute the orbits using $\rho_{out, l}$ because applying OrbitBN to $z_l$ operates on the layer's output space. Using $\rho_{in, l}$ would be both a dimensional and representational mismatch as $\rho_{in, l}$ only constrains symmetries of the previous layer's output, $z_{l-1}$.

All positions related by the transformation belong to the same orbit and must therefore receive identical treatment to preserve equivariance. To apply Orbit-based BN, the mean $\mu$ and variance $\sigma^2$ are computed separately for feature indices per orbit by averaging only over the batch elements and the positions belonging to that orbit. The normalized value, at layer $l$, at each position $i$ then becomes
\begin{equation*}
    \hat{z}_i = \frac{z_i - \mu_{\text{orbit}(i)}}{\sqrt{\sigma^2_{\text{orbit}(i)} + \epsilon}}\;\;\;\epsilon \ll 1,
\end{equation*}
where $\text{orbit}(i)$ represents the orbit to which feature index $i$ belongs to (e.g., if feature indices $j, k$ are in the same orbit then $\text{orbit}(j)=\text{orbit}(k)$ and they share the same shift/scale parameters). A single learned scale parameter $\gamma_{\text{orbit}(i)}$ and shift parameter $\beta_{\text{orbit}(i)}$ (one pair per orbit) are then applied
\begin{equation}\label{Eq: OrbitBN}
    y_i = \gamma_{\text{orbit}(i)} \hat{z}_i + \beta_{\text{orbit}(i)},
\end{equation}
where $y_i$ denotes the normalized feature vector of layer $l$ shifted and scaled using per-orbit parameters. By construction, the OrbitBN method is constant across entire orbits and the application of group transformations (multiplication by their matrix representations) is equivariant: applying OrbitBN before or after multiplication by the group representation does not affect the result (\ref{Eq: OrbitBN}).

Dropout \cite{JMLR:v15:srivastava14a} is a commonly used regularization technique designed to stop networks from overfitting the training data. In a fully connected layer, neurons can develop co-dependencies with other neurons and can learn to make-up for mistakes of their neighbouring neurons rather than actually learning meaningful features. To counteract this, dropout randomly deactivates a percentage of the neurons during every forward pass during training by multiplying them by zero (masking these neurons). This forces the remaining neurons to ``take on more responsibility" and prevents the network from relying heavily on specific connections. Dropout effectively trains an ensemble of exponentially  many thinned networks that share weights. During model testing the full (unthinned) network makes predictions.

However, in the context of equivariance, the conventional dropout method is not suitable. Dropout breaks equivariance as it applies a random mask to the weights with no regard to how features relate to one another. It is therefore possible that during training some features in an orbit are masked whilst the remaining are trained. But for equivariance, all features in an orbit need to share one parameter. By applying dropout randomly, the method fundamentally breaks the needed model symmetry.

This is easily solved by modifying dropout to randomly mask entire orbits, using the orbits of the output representation, rather than the individual features. Following the same logic as with the OrbitBN method, we define dropout on orbits (OrbitDropout) of the output representation ($\rho_{out, l}$) of a layer, $l$. As seen in (\ref{Eq: Activation constraints}) the activations $\sigma_l$ ($l=1,2$) are equivariant with respect to the output representations of layer 1 and 2, respectively. Thus, because Dropout is applied after these activations, it must be defined over the orbits of $ \rho_{out, l}$ to satisfy:
\begin{equation*}
    \rho_{out, l}\text{OrbitDropout}(\sigma_l(z_l)) = \text{OrbitDropout}(\rho_{out, l} \sigma_l(z_l)).
\end{equation*}

Therefore, instead of masking individual and potentially orbit-related features, the orbit-based dropout selects a small percentage of orbits and masks all the features contained within these. This ensures that training the remaining neurons does not interfere with the parameter ties imposed by the equivariance constraint. We perform Orbit-based Dropout after the first two activations ($\sigma_1, \sigma_2$) in layers 1 and 2 with corresponding output representations, $\rho_{out, l} = \varrho_{C_k, l=1}$ and $\rho_{C_k, l=2}$, only.

\subsubsection{Comments on Orbit-regularization}

OrbitBN and OrbitDropout are therefore just altered forms of the classic batch normalization and dropout methods, which are widely used in neural networks. Instead of processing features of a layer independently, which would break equivariance constraints, the two methods presented here operate over orbits, which contain related features. These orbits are obtained by the output representations, $\rho_{out}$, on the indices of a vector $z\in\mathbb{R}^L$ of a given layer:
\begin{equation*}
    \mathcal{O}_{i} = \{\tau^p(i): p=1, \hdots, P\},
\end{equation*}
where $\tau: \{  1, \hdots, L\} \to \{  1, \hdots, L\}$ is a permutation on the indices corresponding to $\rho_{out}$, with $i\in \{  1, \hdots, L\}$ and $\tau^P(i) = i$.\\

Using these orbit-based regularization techniques, we obtain the results for the equivariant networks in tables \ref{tab:ck-results}, \ref{tab:equiv models equiv check}, and \ref{tab:equiv models on defects}, that evidence the utility and success of implementing these functions.

\section{Training and Results}

We discuss now the training and results of the seven networks equivariant to the seven cyclic groups $C_4,\,C_8,\,C_{16},\,C_{32},\,C_{64},\,C_{128}$ and $C_{256}$, and their non-equivariant counter parts (without and with data augmentation). Additionally, we confirm perfect equivariance of the $C_k$-models and how this property does not hold (perfectly) for the non-equivariant networks. We lastly discuss model generalization to completely new defect-type textures.

\subsection{Equivariant Networks}

We present the results of the equivariant models in table \ref{tab:ck-results}. All network results are given as an average $\pm$ sample standard deviation across four independent runs (seed $= 314,\;315,\;316,\;317$). All network architectures input row-wise vectorized black and white images (represented by 0s and 1s) and consist of three hidden layers with three learnable matrices $W_1, W_2$ and $W_3$. The matrix $\mathfrak{L}$ (\ref{General Intertwiner L}) is a non-trainable, predefined matrix mapping from the final hidden layer to the output. Equivariance is implemented by a combination of constraining the weights and choosing specific activations. Each weight matrix is constrained by a specific set of representation matrices. For $W_1$ these are given in table \ref{tab:W1-constraints}, and for $W_2, W_3$ given in table \ref{tab:W2,W3-constraints without L}. The activations used in all networks are GELU (\ref{sigma1, sigma2}) and a norm-based (linear) activation (\ref{sigma3}). The model architectures (layer sizes) and total parameter counts are given in the Appendix in table \ref{tab:ck-architectures}.

The results are given for the best models where $C_4, C_8$ and $C_{16}$ were trained for 25 epochs and $C_{32}, C_{64}, C_{128}$ and $C_{256}$ converged quicker and were trained for only 10 epochs. The learning rate for all equivariant networks was fixed to $0.001$ and batch size for all at 32 (aside from $C_{64}, C_{128}$ and $C_{256}$ which had a batch size of 64 to speed up training).

\begin{table}[h]
\centering
\caption{Equivariant model results averaged  across four different runs.}
\label{tab:ck-results}
\begin{ruledtabular}
\begin{tabular}{ccc}
$C_k$-Model & RMSE$_{Q_{11}}$ & RMSE$_{Q_{12}}$  \\
\hline
$C_{4}$ & 0.0741 $\pm$ 0.0016 & 0.0688 $\pm$ 0.0025  \\
$C_{8}$ & 0.0613 $\pm$ 0.0017 & 0.0600 $\pm$ 0.0019  \\
$C_{16}$ & 0.0567 $\pm$ 0.0022 & 0.0549 $\pm$ 0.0023  \\
$C_{32}$ & 0.0521 $\pm$ 0.0013 & 0.0533 $\pm$ 0.0020 \\
$C_{64}$ & 0.0528 $\pm$ 0.0017 & 0.0544 $\pm$ 0.0009  \\
$C_{128}$ & 0.0450 $\pm$ 0.0012 & 0.0444 $\pm$ 0.0025  \\
$C_{256}$ & 0.0450 $\pm$ 0.0019 & 0.0447 $\pm$ 0.0030  \\
\end{tabular}
\end{ruledtabular}
\end{table}

Equivariant networks successfully learn the relationship between molecule textures and $Q$-values. Increasing the cyclic group order improves generalization and reduces prediction errors, demonstrating that built-in architectural equivariance is far more crucial to model performance than raw parameter count. {For example, highly equivariant models with fewer parameters frequently match or outperform lower-order models with vastly more parameters (such as the $C_{32}$-model vs the $C_4$-model, which differ by nearly one million parameters)}. Overall, the models' Root Mean Square Error (RMSE), which directly translates to the average percentage error across the $Q$-value range (since $Q_{11},\,Q_{12}\in [-0.5,\,0.5]$), remains largely consistent across different texture types, with only the lowest-order model showing slight variations.

Although table \ref{tab:ck-results} gives the RMSE over the whole range of $Q$-values, we find that these approximately represent the RMSE for instances of isotropic $Q$; when we split and compute the individual RMSE by $\lambda_+ = \sqrt{Q_{11}^2 + Q_{12}^2}$ (positive eigenvalue of the $Q$-tensor matrix) when $\lambda_+< 0.1$. Only the $C_4$-model appears to have a slightly lower RMSE for the isotropic cases ($Q_{11}, Q_{12}\sim0.058$) and marginally higher RMSE for the remaining textures ($Q_{11},\,Q_{12}\sim 0.081,\,0.073$).

\subsection{Benchmark Networks}

We present the results of the non-equivariant models without data augmentation in table \ref{tab:non-ck-results} and with augmentation in table \ref{tab:non-ck-results augmented}. All network results are given as an average $\pm$ sample standard deviation across four independent runs (seed $= 314,\;315,\;316,\;317$), similar to the equivariant models. All network architectures input row-wise vectorized black and white images (represented by 0s and 1s) and consists of three hidden layers with three learnable matrices $W_1, W_2$ and $W_3$.As before, the matrix $\mathfrak{L}$ (\ref{General Intertwiner L}) is a non-trainable, predefined matrix mapping from the final hidden layer to the output.

No equivariance is implemented, and thus, the weights are unconstrained and matched approximately to the equivariant networks. Non-equivariant Regularization methods are also used in place of their equivariant orbit-based counterparts, as defined above. We use the same activations as the equivariant models. These choices give us approximately equivalent benchmarks to compare the equivariant networks with.

The results are given for the best models where all networks are trained for 25 epochs (aside from MLP$_{C_{128}}$ and MLP$_{C_{256}}$ which were trained for 50 and 100 epochs, respectively). The learning rate for all networks presented was set to 0.001 and batch size was fixed for all at 32.

\begin{table}[h]
\centering
\caption{Non-equivariant model results. No data-augmentation used during training.}
\label{tab:non-ck-results}
\begin{ruledtabular}
\begin{tabular}{ccc}
Non-aug. MLP$_{C_k}$-Model &  RMSE$_{Q_{11}}$ & RMSE$_{Q_{12}}$  \\
\hline
MLP$_{C_{4}}$ & 0.0994 $\pm$ 0.0039 & 0.1323 $\pm$ 0.0056  \\
MLP$_{C_{8}}$ & 0.0963 $\pm$ 0.0053 & 0.0992 $\pm$ 0.0041  \\
MLP$_{C_{16}}$ & 0.1041 $\pm$ 0.0011 & 0.0979 $\pm$ 0.0063  \\
MLP$_{C_{32}}$ & 0.1038 $\pm$ 0.0026 & 0.0976 $\pm$ 0.0103  \\
MLP$_{C_{64}}$ & 0.1061 $\pm$ 0.0067 & 0.0981 $\pm$ 0.0108  \\
MLP$_{C_{128}}$ & 0.0977 $\pm$ 0.0013 & 0.0987 $\pm$ 0.0021  \\
MLP$_{C_{256}}$ & 0.2635 $\pm$ 0.0059 & 0.2508 $\pm$ 0.0081   
\end{tabular}
\end{ruledtabular}
\end{table}

The results of the non-augmented non-equivariant models, whilst not purely random, are overall less accurate than their symmetric counterparts. Even the lowest RMSE of the non-augmented MLP$_{C_k}$ models (MLP$_{C_{8}}$ where RMSE$_{Q_{11}} \approx 0.09$ and RMSE$_{Q_{12}}\approx 0.09$) are higher than the worst errors of the $C_k$-equivariant networks ($C_4$ where where RMSE$_{Q_{11}} \approx 0.07$ and RMSE$_{Q_{12}}\approx 0.06$). In an attempt to improve these models, we employ data augmentation, and observe the results.


It is well known that models can benefit from augmenting the training set \cite{10.1145/3065386}, particularly when the initial data set is small. The augmentations usually take the form of rotations, scaling, or other transformations that inject noise or fuzziness into the training set. Data labels are transformed accordingly or remain invariant to these modifications.

In our approach to data augmentation we require that when an image is transformed that its label is modified according to the known equivariance relation of the $Q$-tensor: rotation of all molecules (with an associated Tensor $Q$) by $\theta$, must transform the original $Q$ to $Q \mapsto Q' =  RQR^T$ where $R$ is a $2\times 2$ counter-clockwise rotation matrix evaluated at $2\theta$. This gives a new data point and label pair ($R$(Image), $Q'$). Note that rotation of individual particles is equivalent to rotating the image containing these. $Q$ is translation invariant and so does not change if particles move.

To avoid storing the augmented images on disk, we augment on-the-fly (during training). More specifically, the data is transformed such that each training instance has a 50\% chance of being augmented via an counter-clockwise rotation by a random angle $\alpha = a \frac{\pi}{k}$, for some $a=1, 2, \hdots, 2k$, using bicubic interpolation. The results of the augmented models are presented in table \ref{tab:non-ck-results augmented}. As expected, all the augmented models outperform their non-augmented counterparts. However, even with a larger training set, they appear to not generalize as well to the test data as the equivariant models do, as given in table \ref{tab:ck-results}.

\begin{table}[h]
\centering
\caption{Non-equivariant model results with data augmentation.}
\label{tab:non-ck-results augmented}
\begin{ruledtabular}
\begin{tabular}{ccc}
Aug. MLP$_{C_k}$-Model &  RMSE$_{Q_{11}}$ & RMSE$_{Q_{12}}$  \\
\hline
MLP$_{C_{4}}$ & 0.0909 $\pm$ 0.0011 & 0.1025 $\pm$ 0.0031  \\
MLP$_{C_{8}}$ & 0.0877 $\pm$ 0.0031 & 0.0904 $\pm$ 0.0013 \\
MLP$_{C_{16}}$ & 0.0916 $\pm$ 0.0021 & 0.0906 $\pm$ 0.0065  \\
MLP$_{C_{32}}$ & 0.0929 $\pm$ 0.0010 & 0.0916 $\pm$ 0.0109  \\
MLP$_{C_{64}}$ & 0.0952 $\pm$ 0.0013 & 0.0929 $\pm$ 0.0079  \\
MLP$_{C_{128}}$ & 0.0834 $\pm$ 0.0078 & 0.0861 $\pm$ 0.0064  \\
MLP$_{C_{256}}$ & 0.1158 $\pm$ 0.0741 & 0.1244 $\pm$ 0.0918  \\
\end{tabular}
\end{ruledtabular}
\end{table}

Data augmentation does appear to improve model performance on the test data across all networks. However, even the lowest RMSEs (given by model MLP$_{C_{128}}$ where RMSE$_{Q_{11}} \approx 0.08$ and RMSE$_{Q_{12}}\approx 0.08$) are higher than the highest errors given by the $C_4$-equivariant model.

It therefore suggests that the addition of equivariance through weight ties into the models does improve model generalization, even when number of parameters are approximately matched and same activations used. The use of data augmentation is still outperformed by the equivariant models.

\subsection{Validation of Equivariance \& Generalization to Defect Data}


We find in table \ref{tab:equiv models equiv check} that the $C_k$-models are all equivariant up to machine precision (floating point 32 used for all models), demonstrating that the $Q$-tensor constraint
\begin{equation}\label{Eq:Perm Test}
    |\nu(\varrho_{C_k, 0}(g)x)_{i}- (R_{\frac{4p\pi}{k}}\nu(x))_{i}| = 0\; \text{for}\; i=1,2
\end{equation}
holds over the four independent runs. The network's prediction on the permuted input $\nu(\varrho_{C_k, 0}(g)x)$ needs to match the transformed output of the models untransformed input $R_{\frac{4p\pi}{k}}\nu(x)$. This is evaluated using the transformed prediction $\nu(\varrho_{C_k, 0}(g)x)_{\text{pred}}$ and the untransformed prediction $\nu(x)_{\text{pred}}$. Alternatively, one could evaluate equivariance using the true values too. As intended, constraining all parameters, activations and regularization techniques to be equivariant results in very low errors (perfect up to machine precision). We give the equivariance errors of the non-equivariant benchmark models in table \ref{tab:non-equiv models equiv test} for comparison.



We additionally evaluate the equivariant and non-equivariant models on data that was not part of the training set \ref{tab:equiv models on defects}. This data takes the form of hedgehog-like molecular arrangements, where the molecules are organized in a point defect fashion. However, their corresponding $Q$-values are close to zero. Of interest is (1) how well the models generalize to totally new data, and (2) whether the networks are able of capturing the relationship between ellipse arrangements and $Q$.

On the textures not seen during training, the lowest errors the $C_k$-equivariant models achieve for both $Q_{11}$ and $Q_{12}$ is around $\sim 0.04$ (for model $C_{256}$) and $\sim 0.07$ (for the $C_{32}$ model). The highest errors are found in the $C_4$-network. In contrast, the MLP$_{C_k}$-models without augmentation generalize quite poorly, especially for MLP$_{C_{256}}$.

Data augmenting the non-equivariant networks does significantly improve the performance, bringing it closer to its equivariant counterpart, and at times outperforming these (such as MLP$_{C_{32}}$, MLP$_{C_{64}}$, MLP$_{C_{128}}$). The results indicate that built-in equivariance or explicit symmetry augmentation is essential for generalization to these defect configurations.

\section{Conclusion}


We presented a method by which we constructed several $C_k$-equivariant neural networks; we trained and evaluated these on clean, synthetically generated microscopic nematic liquid crystal textures. To do this, we constructed $C_k$-representations that act as approximate rotations of circular subdomains of images of nematic textures. Using these and the regular representations of the cyclic groups, we built neural networks which satisfy the rotational equivariance of the $Q$-tensor exactly (up to machine precision). Comparing these symmetric networks to benchmarks models, we find that they outperform these benchmarks on several fronts: they yield lower RMSE errors, are mathematically equivariant, and are more robust to new (defect-type) data.

One limitation of this work is perhaps the use of synthetic training data; the use of more realistic data could make these equivariant networks more applicable for real-world uses, such as in the study of thin-film nematic liquid crystals. We additionally limit ourselves to the two-dimensional model, and do not address the $Q$-tensor order parameter $Q\in\mathbb{R}^{3\times 3}$ which describes structure in three-dimensional nematic textures. Other types of symmetries in liquid crystals (relevant to cholesterics or smectics) could be studied by modifying the order parameter and activation constraints as needed.

In summary, this work demonstrates that in our case encoding a known bias into neural network architectures proves to be a useful choice over learning equivariance constraints manually (such as through data augmentation). Our Python codes which were used to generate the results in this work can be found on \href{https://github.com/NavarroJulia/On-the-Equivariant-Learning-of-the-Q--Tensor-Order-Parameter}{GitHub}.

\bibliographystyle{amsplain}
\bibliography{bibby}

\appendix

\clearpage

\onecolumngrid

\section{Model Tables}

\subsection*{Model Architectures}

\begin{table}[h]
\centering
\caption{Model architectures equivariant to different cyclic groups $C_k$. }
\label{tab:ck-architectures}
\begin{ruledtabular}
\begin{tabular}{cccc}
 Equiv. to Group $C_k$ & Model architecture  & Total trainable par. (incl. O-BN par.) & Reduction factor \\
\hline
$C_{4}$ & $250^2 \xrightarrow{W_1} 100 \xrightarrow{W_2} 8 \xrightarrow{W_3} 8 \xrightarrow{\mathfrak{L}} 2$ 
     & {1,562,774} & 4.00$\times$  \\
$C_{8}$ & $250^2 \xrightarrow{W_1} 100 \xrightarrow{W_2} 8 \xrightarrow{W_3} 4 \xrightarrow{\mathfrak{L}} 2$ 
     & {1,215,768} & 5.14$\times$  \\
$C_{16}$ & $250^2 \xrightarrow{W_1} 100 \xrightarrow{W_2} 16 \xrightarrow{W_3} 8 \xrightarrow{\mathfrak{L}} 2$ 
   & {856,432}  & 7.30$\times$  \\
$C_{32}$ & $250^2 \xrightarrow{W_1} 100 \xrightarrow{W_2} 32 \xrightarrow{W_3} 16 \xrightarrow{\mathfrak{L}} 2$ 
   & {676,644}  & 9.24$\times$  \\
$C_{64}$ & $250^2 \xrightarrow{W_1} 100 \xrightarrow{W_2} 64 \xrightarrow{W_3} 32 \xrightarrow{\mathfrak{L}} 2$ 
   & {586,762}  & 10.67$\times$  \\
$C_{128}$ & $250^2 \xrightarrow{W_1} 196 \xrightarrow{W_2} 128 \xrightarrow{W_3} 64 \xrightarrow{\mathfrak{L}} 2$ 
   & {1,031,222} & 11.91$\times$  \\
$C_{256}$ & $250^2 \xrightarrow{W_1} 400 \xrightarrow{W_2} 256 \xrightarrow{W_3} 128 \xrightarrow{\mathfrak{L}} 2$ 
   & {2,086,522} & 12.05$\times$  
\end{tabular}
\end{ruledtabular}
\end{table}

The $\times$-fold parameter reduction is computed for the equivariant (weight-sharing/constrained) models relative to their non-equivariant (unconstrained, no weight tying) counterparts, considering only the trainable parameters. The first-layer output channel size varies across models as we use our custom permutation matrix in the equivariance constraint on $W_1$.



\begin{table}[h]
\centering
\caption{Model architectures for non-equivariant models and their total trainable parameter count.}
\label{tab:non-ck-architectures}
\begin{ruledtabular}
\begin{tabular}{ccc}
Model Name & Model architecture & Total trainable par.   \\
\hline
MLP$_{C_{4}}$ &  \;$250^2 \xrightarrow{W_1} 25 \xrightarrow{W_2} 8 \xrightarrow{W_3} 8 \xrightarrow{\mathfrak{L}} 2$ 
 &   1,562,764\\
MLP$_{C_{8}}$ &  \;$250^2 \xrightarrow{W_1} 19 \xrightarrow{W_2} 8 \xrightarrow{W_3} 4 \xrightarrow{\mathfrak{L}} 2$ 
 &   1,187,684\\
MLP$_{C_{16}}$ &  \;$250^2 \xrightarrow{W_1} 14 \xrightarrow{W_2} 16 \xrightarrow{W_3} 8 \xrightarrow{\mathfrak{L}} 2$ 
 &  875,352\\
MLP$_{C_{32}}$ &  \;$250^2 \xrightarrow{W_1} 11 \xrightarrow{W_2} 32 \xrightarrow{W_3} 16 \xrightarrow{\mathfrak{L}} 2$ 
 &   688,364\\
MLP$_{C_{64}}$ &  \;$250^2 \xrightarrow{W_1} 9 \xrightarrow{W_2} 64 \xrightarrow{W_3} 32 \xrightarrow{\mathfrak{L}} 2$ 
 &  565,124\\
MLP$_{C_{128}}$ &  \;$250^2 \xrightarrow{W_1} 16 \xrightarrow{W_2} 128 \xrightarrow{W_3} 64 \xrightarrow{\mathfrak{L}} 2$ 
 &  1,010,240 \\
MLP$_{C_{256}}$ &  \;$250^2 \xrightarrow{W_1} 33 \xrightarrow{W_2} 256 \xrightarrow{W_3} 128 \xrightarrow{\mathfrak{L}} 2$ 
 &  2,103,716 
\end{tabular}
\end{ruledtabular}
\end{table}

 The architecture is constructed so parameter counts are matched, giving the MLP$_{C_k}$ models approximately the same number of parameters as the $C_k$-models. In an attempt to preserve the architecture of the equivariant models (so they are approximately the same for the MLP$_{C_k}$-networks), we only modify the size of the first layer and keep the non-trainable matrix $\mathfrak{L}$.

\clearpage


\subsection*{Equivariance of $C_k$-models}

\begin{table}[h]
\caption{Model equivariance verified for both $Q_{11}$ and $Q_{12}$.}
\label{tab:equiv models equiv check}
\begin{ruledtabular}
\begin{tabular}{lcc}
 & \multicolumn{2}{c}{Check that $|\nu(\varrho_{C_k, 0}(g)x)_{i}- )R_{\frac{4p\pi}{k}}\nu(x))_{i}|$ = 0 for $i=1,2$}  \\ \cline{2-3}
$C_k$-Model &
  RMSE$_{Q_{11}}$  & RMSE$_{Q_{12}}$ \\ \hline
$C_{4}$ & 0.000001709 $\pm$ 0.000000239  &  0.000001787 $\pm$ 0.000000231      \\
$C_{8}$   & 0.000001079 $\pm$ 0.000000103 &  0.000001110 $\pm$ 0.000000061  \\
$C_{16}$  & 0.000001185 $\pm$ 0.000000051 &  0.000001216 $\pm$ 0.000000048 \\
$C_{32}$  & 0.000001294 $\pm$ 0.000000076 & 0.000001289 $\pm$ 0.000000074  \\
$C_{64}$  & 0.000001410 $\pm$ 0.000000208 &  0.000001371 $\pm$ 0.000000153 \\
$C_{128}$ & 0.000000992 $\pm$ 0.000000192 &  0.000001018 $\pm$ 0.000000252 \\
$C_{256}$ & 0.000000933 $\pm$ 0.000000245 &  0.000000985 $\pm$ 0.000000191 \\
\end{tabular}
\end{ruledtabular}
\end{table}

For each architecture we report the mean RMSE $\pm$ sample standard deviation over four independent runs (different seeds). All networks were trained and evaluated in single precision (fp32) and thus the observed errors ($\approx 1-2 \times 10^{-6}$) are at the level of accumulated round-off error. Therefore, within single-precision numerical accuracy the models can be considered perfectly equivariant. It is sufficient to test for $g=g^1$ as these correspond to the generator elements of the cyclic group; if equivariance holds for $g$, it holds for all powers of $g$.



\subsection*{Equivariance of MLP$_{C_k}$-models}

\begin{table}[h]
\caption{Equivariance errors of the non-equivariant models (without and with data augmentation), checking if the equivariance condition holds for the various models. Data augmentation improves model equivariance, however, they are still on the order of $10^{-2}$ (roughly 4 order of magnitudes worse than the equivariant models).}
\label{tab:non-equiv models equiv test}
\begin{ruledtabular}
\begin{tabular}{lcccc}
 & \multicolumn{4}{c}{Check that $|\nu(\varrho_{C_k, 0}(g)x)_{i}- (R_{\frac{4p\pi}{k}}\nu(x))_{i}|$ = 0 for $i=1,2$} 
 \\ \cline{2-5} 
 
 & \multicolumn{2}{c}{No Data Augmentation}
 & \multicolumn{2}{c}{With Data Augmentation}
 \\ \cline{2-3} \cline{4-5} 
  MLP$_{C_k}$-Model &
  RMSE$_{Q_{11}}$ & RMSE$_{Q_{12}}$ &
  RMSE$_{Q_{11}}$ & RMSE$_{Q_{12}}$
  \\ \hline
 MLP$_{C_{4}}$ &  0.0791  $\pm$ 0.0087 &  0.1129  $\pm$ 0.0219  & 0.0495 $\pm$ 0.0058 &  0.1169  $\pm$ 0.0172  \\ 

 MLP$_{C_{8}}$ & 0.0786 $\pm$ 0.0073 &  0.0866  $\pm$ 0.0098  &  0.0591  $\pm$ 0.0054  &  0.0610  $\pm$ 0.0032  \\ 

 MLP$_{C_{16}}$ & 0.0674  $\pm$ 0.0021 &  0.0626  $\pm$ 0.0096  & 0.0541   $\pm$ 0.0061  &  0.0494  $\pm$ 0.0043  \\ 

MLP$_{C_{32}}$ & 0.0621  $\pm$ 0.0097 &  0.0565  $\pm$ 0.0132  & 0.0478  $\pm$ 0.0062 &  0.0450  $\pm$ 0.0043  \\ 

MLP$_{C_{64}}$ & 0.0511   $\pm$ 0.0077 &  0.0526  $\pm$ 0.0011 & 0.0387  $\pm$  0.0036 &  0.0425  $\pm$ 0.0030 \\ 

 MLP$_{C_{128}}$ & 0.0548  $\pm$ 0.0072 &  0.0543  $\pm$ 0.0060 & 0.0520  $\pm$ 0.0066 &   0.0467  $\pm$ 0.0051 \\ 

 MLP$_{C_{256}}$ & 0.1998  $\pm$ 0.0495 &  0.1799  $\pm$ 0.0424 & 0.0821  $\pm$ 0.0835 &  0.0741  $\pm$ 0.0710  \\ 
\end{tabular}
\end{ruledtabular}
\end{table}

\newpage


\subsection*{Model performance on Defect data}

\begin{figure}[h]
    \centering
    
    \begin{subfigure}{0.20\textwidth}
        \centering
        \includegraphics[width=\linewidth]{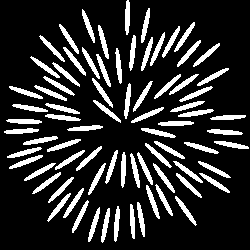}
        \caption{}
    \end{subfigure}
    \hspace{1cm}
    \begin{subfigure}{0.20\textwidth}
        \centering
        \includegraphics[width=\linewidth]{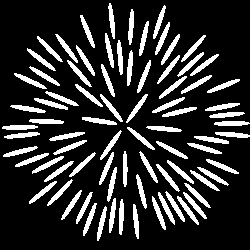}
        \caption{}
    \end{subfigure}
    \hspace{1cm}
    \begin{subfigure}{0.20\textwidth}
        \centering
        \includegraphics[width=\linewidth]{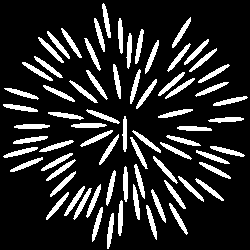}
        \caption{}
    \end{subfigure}
    \caption{Defect Textures; appear ordered, however correspond to small $Q$-values. Not part of the original 50,000-image dataset.}
    \label{fig:Defects}
\end{figure}

\begin{table}[h]
\caption{Model performance on the hedgehog/defect data (see Figure \ref{fig:Defects}). This type of data is completely new to the models; it was not used during the learning stage. The images appear ordered (ellipse particles arranged in circles pointing towards the centre), however the associated $Q$-tensor values are nearly isotropic (close to zero).}
\label{tab:equiv models on defects}
\begin{ruledtabular}
\begin{tabular}{lccccccc}
 & \multicolumn{2}{c}{}
 & \multicolumn{1}{c}{}
 & \multicolumn{2}{c}{No Data Augmentation}
 & \multicolumn{2}{c}{With Data Augmentation}
 \\ \cline{5-6}\cline{7-8}
$C_k$-Model &
  RMSE$_{Q_{11}}$ & RMSE$_{Q_{12}}$ &
  MLP$_{C_k}$-Model &
  RMSE$_{Q_{11}}$ & RMSE$_{Q_{12}}$ &
  RMSE$_{Q_{11}}$ & RMSE$_{Q_{12}}$
  \\ \hline
$C_{4}$   &  0.1382 $\pm$ 0.0090 &  {\color{gray}0.1526 $\pm$ 0.0031}  & MLP$_{C_{4}}$ &  0.1768 $\pm$ 0.0095 & {\color{gray}0.2018 $\pm$ 0.0035} & 0.1764 $\pm$ 0.0142 & {\color{gray}0.2137  $\pm$ 0.0050} \\ 

$C_{8}$   &  0.0504 $\pm$ 0.0030 &  {\color{gray}0.0531 $\pm$ 0.0036}  & MLP$_{C_{8}}$ & 0.0715 $\pm$ 0.0061 & {\color{gray}0.0650 $\pm$ 0.0080} &  0.0517 $\pm$ 0.0066 & {\color{gray}0.0533 $\pm$ 0.0034} \\ 

$C_{16}$  &  0.0380 $\pm$ 0.0031 &  {\color{gray}0.0389 $\pm$ 0.0023}  & MLP$_{C_{16}}$ & 0.0725 $\pm$ 0.0132 & {\color{gray}0.0860 $\pm$ 0.0239} & 0.0451 $\pm$ 0.0072 & {\color{gray}0.0453 $\pm$  0.0077} \\ 

$C_{32}$  &  0.0664 $\pm$ 0.0203 &  {\color{gray}0.0699 $\pm$ 0.0246}  & MLP$_{C_{32}}$ & 0.1278 $\pm$ 0.0629 & {\color{gray}0.1119 $\pm$ 0.0449} & 0.0464 $\pm$ 0.0105 & {\color{gray}0.0543  $\pm$ 0.0152} \\ 

$C_{64}$  &  0.0632 $\pm$ 0.0272  &  {\color{gray}0.0642 $\pm$ 0.0284}  & MLP$_{C_{64}}$ & 0.1909 $\pm$ 0.2032 & {\color{gray}0.0833 $\pm$ 0.0124} & 0.0443 $\pm$ 0.0151  & {\color{gray}0.0383 $\pm$ 0.0041} \\ 

$C_{128}$ &  0.0654 $\pm$ 0.0194 & {\color{gray}0.0630 $\pm$ 0.0182}  & MLP$_{C_{128}}$ & 0.1341 $\pm$ 0.0309 & {\color{gray}0.1318 $\pm$ 0.0274} & 0.0426 $\pm$ 0.0051 &  {\color{gray}0.0394 $\pm$ 0.0029} \\ 

$C_{256}$ &  0.0409 $\pm$ 0.0155 &  {\color{gray}0.0406 $\pm$ 0.0152}  & MLP$_{C_{256}}$ & 0.3520 $\pm$ 0.0036 & {\color{gray}0.3432 $\pm$ 0.0042} & 0.1869 $\pm$ 0.1169 & {\color{gray}0.1750 $\pm$  0.1084} \\ 
\end{tabular}
\end{ruledtabular}
\end{table}

\section{Runtime}\label{Runtime}

The reported times given in table \ref{tab:runtimes} are measured from the creation of the configuration file (precedes initialization and training) to the saving of the last optimal model checkpoint (based on validation performance) and until the final model (reached at final epoch). Time is given in mins.secs as an average $\pm$ sample standard deviation over four different seed runs (314, 315, 316, 317). The Total Epochs column acts as a comparison across the different network architectures.

\begin{table}[h]
\caption{Runtimes of all models across four different runs.}
\label{tab:runtimes}
\begin{ruledtabular}
\begin{tabular}{lccc}
\textrm{Model} &
\textrm{Total Epochs} &
\textrm{Time (mins.secs) best model} & \textrm{Time (mins.secs) final model} \\
\\
\colrule
\multicolumn{4}{l}{$C_k$-equivariant models} \\
\colrule
$C_{4}$  & 25 & 78.17 $\pm$ 14.31 & 86.01 $\pm$ 0.36   \\
$C_{8}$  & 25 & 60.09 $\pm$ 9.15 & 101.22 $\pm$ 0.38   \\
$C_{16}$  & 25 & 35.34 $\pm$ 5.02 & 94.01 $\pm$ 0.38   \\
$C_{32}$  & 10 & 56.48 $\pm$ 12.07 & 71.09 $\pm$ 9.29   \\
$C_{64}$  & 10 & 37.10 $\pm$ 20.19 & 42.18 $\pm$ 22.15   \\
$C_{128}$  & 10 & 54.23 $\pm$ 24.03 & 81.44 $\pm$ 44.58   \\
$C_{256}$  & 10 & 80.46 $\pm$ 18.13 & 106.10 $\pm$ 30.32   \\
\colrule
\multicolumn{4}{l}{Non-augmented non-equivariant models} \\
\colrule
MLP$_{C_{4}}$ & 25 & 2.42 $\pm$ 0.19 & 3.21 $\pm$ 0.09   \\
MLP$_{C_{8}}$ & 25 & 1.55 $\pm$ 0.57 & 3.32 $\pm$ 0.16    \\
MLP$_{C_{16}}$  & 25 & 1.05 $\pm$ 0.40 & 3.28 $\pm$ 0.07    \\
MLP$_{C_{32}}$  & 25 & 0.54 $\pm$ 0.44 & 3.30 $\pm$ 0.10   \\
MLP$_{C_{64}}$  & 25 & 1.10 $\pm$ 0.51 & 3.34 $\pm$ 0.26   \\
MLP$_{C_{128}}$  & 50 & 3.20 $\pm$ 0.33 & 6.52 $\pm$ 0.34   \\
MLP$_{C_{256}}$  & 100 & 9.51 $\pm$ 2.57 & 13.23 $\pm$ 1.07    \\
\colrule
\multicolumn{4}{l}{Augmented non-equivariant models} \\
\colrule
MLP$_{C_{4}}$  & 25 & 4.19 $\pm$ 0.20 & 4.38 $\pm$ 0.15  \\
MLP$_{C_{8}}$  & 25 & 6.34 $\pm$ 0.08 & 7.30 $\pm$ 0.24   \\
MLP$_{C_{16}}$  & 25 & 7.13 $\pm$ 0.50 & 8.07 $\pm$ 0.24  \\
MLP$_{C_{32}}$  & 25 & 7.33 $\pm$ 1.13 & 8.16 $\pm$ 0.06  \\
MLP$_{C_{64}}$  & 25 & 7.52 $\pm$ 0.82 & 8.33 $\pm$ 0.06  \\
MLP$_{C_{128}}$  & 50 & 16.53 $\pm$ 0.35 & 17.36 $\pm$ 0.25   \\
MLP$_{C_{256}}$  & 100 & 33.18 $\pm$ 0.51 & 34.09 $\pm$ 0.34  \\
\end{tabular}
\end{ruledtabular}
\end{table}

In contrast with the equivariant networks, the times to reach an optimal model are shorter and there is less variability across these times for the non-equivariant models, even with approximate parameter counts across architectures. However, none of the equivariant best model training times exceeded 100 minutes, demonstrating that whilst training might take longer, it gives the guarantee of perfect equivariance, unlike traditional neural networks.

Training was conducted on a workstation equipped with four NVIDIA RTX A6000 48GB GPUs (48 GB VRAM each, 192 GB total), an Intel Xeon Gold 6230 CPU (80 cores), and 810 GB of system RAM, running Python 3.8.5 with PyTorch on CUDA 8.6.


\newpage

\section{Construction of the Rotation-like Permutation Matrix for $C_k$ Groups of any Order}\label{Construction of the Rotation-like Permutation Matrix}

\subsection{Construction Process}

A representation $\varrho_{C_k}(g) : C_k \to GL(N^2)$ is constructed by first defining the representation of the generator element $g\in C_k$, and then extending these to the remaining elements by $\varrho_{C_k}(g^p)=\varrho_{C_k}(g)^p$. We follow these 4 steps:

\begin{itemize}
    \item Partition the image into concentric rings and fixed points
    \item Impose angular ordering on the pixels within each ring
    \item Define a cyclic permutation on each ring of order $k$, and assemble these, and the fixed points, into a global matrix $\varrho_{C_k}(g)$ acting on {row-wise vectorized} images
    \item The matrices corresponding to the remaining elements $g^2, ..., g^k \in C_k$ are obtained by $\varrho_{C_k}(g^p) = \varrho_{C_k}(g)^p$
\end{itemize}

These steps are discussed in more detail next. For simplicity, we refer $\varrho_{C_k}(g)$ and $\varrho(g)$.
\subsubsection{Partitioning of Images into Rings}



Pixels coordinates $(i,j)$ are organized such that $(0,0)$ is the top-left, such that $i$ increases downward and $j$ increases to the right, where $0\leq i,j < N$ with centre $c=\lfloor \frac{N}{2}\rfloor$. The radius of a pixel $(i,j)$ is
\begin{equation*}
    r(i,j) =\text{round}(\sqrt{(i-c)^2 + (j-c)^2})
\end{equation*}
where we round to the nearest integer. The algorithm groups pixels by their radii (only those with radii $\leq \frac{N}{2}$). Pixels with the same integer radius form a ring. This partitions the grid into into multiple rings (sets) $R = R_1, R_2 \ldots , R_{\text{max}}$, excluding the centre pixel $R_0$ which remains fixed. Each $R$ corresponds to a set of pixels which share the same integer radius.

These rings are then progressively merged as we require the rings to have a multiple of $k$ pixels within them. This begins from the innermost ring $R_1$, where pixels from consecutive rings ($R_2,\;R_3,\;...$) are combined with $R_1$'s pixels until the combined set has a size ($m$ pixels) divisible by $k$, at which point it forms a single merged ring ($\tilde{R_1},\;\tilde{R_2},\;...$) on which permutation can be implemented.

However, it should be noted that not all pixels in the circular region can necessarily be incorporated into perfect rings; if the total number of pixels available cannot be perfectly partitioned, the remaining pixels (typically from the outermost rings that do not accumulate to a multiple of $k$) are treated as fixed points. The algorithm guarantees that every formed ring has size divisible by $k$, but cannot guarantee that all pixels can be incorporated into rings. The remaining pixels that cannot be incorporated into rings with pixel count divisible by $k$ are treated as fixed points and typically occur near the boundary of the circular region.

\subsubsection{Angular Ordering}

In order to implement a rotation-like permutation on pixels in a ring we require a notion of angular ordering. For each pixel in a ring 
\begin{equation*}
    \tilde{R}=\tilde{R_1},\;\tilde{R_2},\;...,\;\tilde{R}_{max}
\end{equation*}
we compute its polar angle with respect to the centre
\begin{equation*}
    \theta(i,j) = \text{atan2}(-(i-c), j-c) \in (-\pi, \pi]
\end{equation*}
Note that $\text{atan2}(y,x)$ returns the counter-clockwise polar angle of a vector from the centre to the pixel location. It is measured from the positive $x$-axis and correctly handles all quadrants. The pixels of each ring $\tilde{R}$ are sorted by increasing values of $\theta$, yielding an ordered list of $m$ pixels (where $\frac{m}{k} \in \mathbb{Z}^{+}$)
%
\begin{equation*}
    \tilde{R}_{ordered}= \{p_0, p_1, \ldots , p_{(m-1)}\}
\end{equation*}

\subsubsection{Cyclic Shift}

The linear transformation $\varrho(g)$ needs to be cyclic, i.e., $\varrho(g)^k=\mathbb{I}$. To enforce this on each ring independently, the algorithm chooses a shift value $s$ (\textbf{per ring}) such that one application of it to a pixel in $\tilde{R}_{ordered}$ moves it forward by $s=\frac{m}{k}$ positions. Since $ks=m$, applying the shift $k$ times yields the identity permutation (pixels return to their original positions).

Since $s$ gives us the value which we shift all pixels $p_a$, where $a=0,\;...,\;m-1$, in each $\tilde{R}_{ordered}$ set, we express it as
\begin{equation*}
    {p}_{a}\to {p}_{(a+s)}
\end{equation*}
and after $k$ applications of applying the shift to pixels $p_a$ we have:
\begin{equation*}
    {p}_{a} \xrightarrow{}{p}_{(a+s)} \xrightarrow{} \hdots \xrightarrow{} {p}_{(a+ks)} = {p}_{(a+m)}\equiv {p}_{a\;(\text{mod}\;m)}
\end{equation*}
The centre pixel, corner pixels (those with radius larger than $\frac{N}{2}$), and any other pixels that are not part of any ring are treated as fixed point and map to themselves
\begin{equation*}
    {p}_{fixed} \xrightarrow{}{p}_{fixed}
\end{equation*}

\subsubsection{Assembling into a Matrix}

These ring permutations can be represented by a single global matrix 
\begin{equation*}
    \varrho(g)\in \{0, 1\}^{N^2\times N^2}
\end{equation*}
representing the map from current pixel locations to their ``rotated" counterparts. The permutation matrix is constructed so that column indices correspond to original (vectorized) pixel positions and row indices correspond to transformed (vectorized) pixel positions:
\begin{equation*}
    \varrho(g)[\text{destination},\;\text{source}] = 1
\end{equation*}
For pixels ${p}$ in each set $\tilde{R}_{ordered}$ with a corresponding shift $s$, we insert its entries into $\varrho(g)$ as follows
\begin{equation*}
    \varrho(g)[i_{(a+s)\text{mod}\;m}N +j_{(a+s)\text{mod}\;m},\;i_{a}N+j_{a}]=1\;\;\text{for all}\;a=0,\ldots , m-1
\end{equation*}
where pixels ${p}_a = (i_a, j_a)$ transform to ${p}_{(a+s)} = (i_{(a+s)}, j_{(a+s)})$. Recall that we are working on the row-wise flattened image, hence the 2D coordinate $(i,j)$ transforms to $iN+j$.

For the fixed points
\begin{equation*}
    \varrho(g)[i_fN+j_f,\;i_fN+j_f] = 1
\end{equation*}
where ${p}_f = (i_f, j_f)$ is a fixed point: centre, outside mask, or unassigned pixel.

Since each pixel maps to exactly one destination and each destination receives exactly one pixel, $\varrho(g)$ is a valid permutation matrix, satisfying $\varrho(g)^k=\mathbb{I}$. Code for this algorithm can be found on \href{https://github.com/NavarroJulia/On-the-Equivariant-Learning-of-the-Q--Tensor-Order-Parameter}{GitHub}. By modifying the code parameters $N$ (image size, assumed to be square) and $k$ (corresponding to the group size of the cyclic group $C_k$; also dictates the ``rotation-angle": $\frac{2\pi}{k}$), one can construct new $\varrho$-type representations and test them on new images.







\section{Dataset Generation}

The data needed to train the models takes the form of pairs of images coupled with their associated $Q$-tensor values. To generate the synthetic dataset needed for training, we employ a stochastic simulation of hard ellipses confined within a circular (or square) domain. Note that we generate the molecular configurations on a square domain exclusively for the $C_4$-equivariant model; the remaining $C_k$-equivariant models are all trained on data where the ellipses are plotted within a circular subdomain. This is related to how the rotation-like matrix representations act on these images: it essentially acts as a rotation by $\frac{2\pi}{k}$ of the inner circular subdomain, and fixes the corners of the main square domain.

\begin{figure}[h] 
\centering
\includegraphics[width=1.0\columnwidth]
{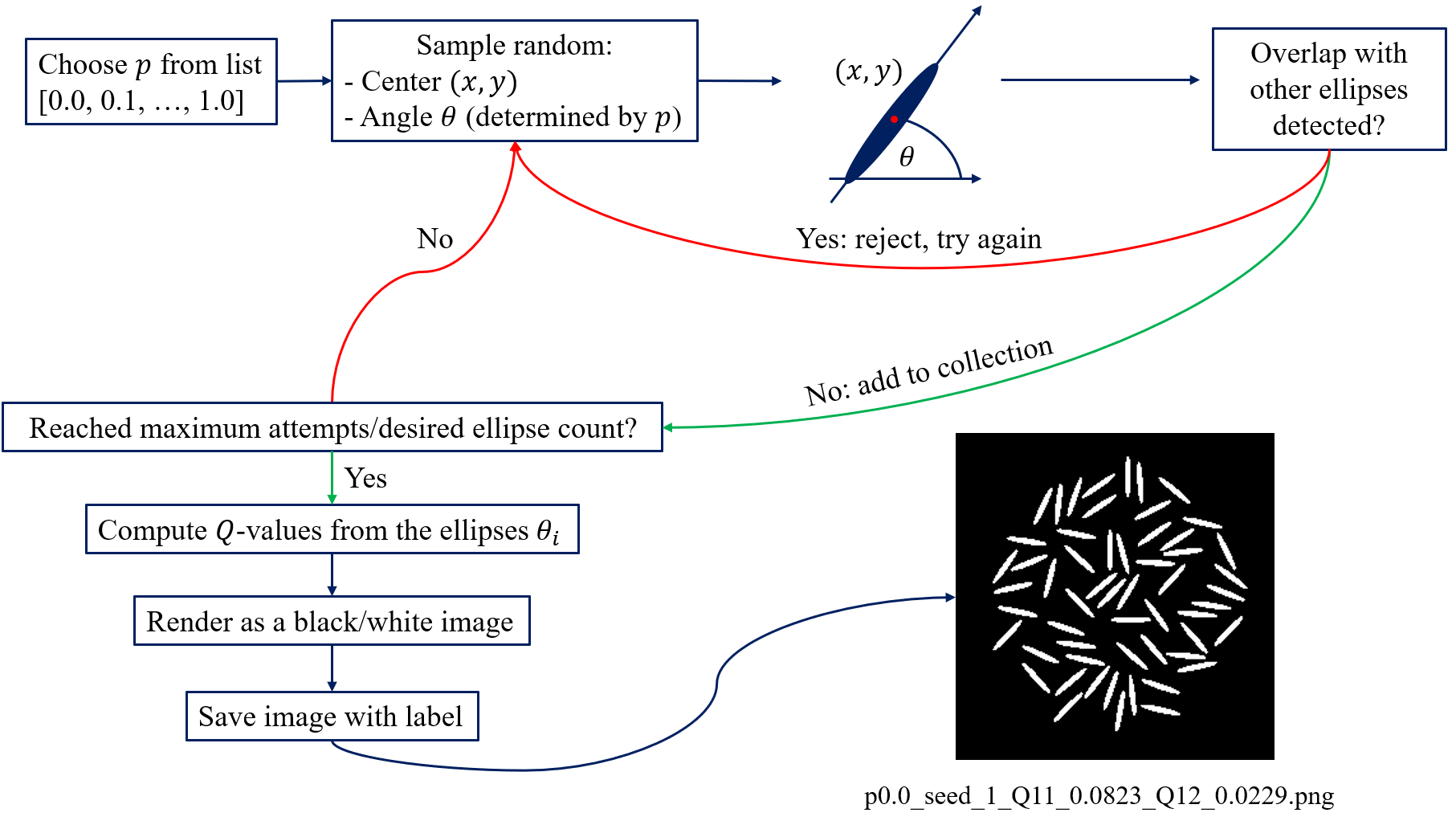}
\caption{{Diagram illustrating the data generation process of a single image: starting off by selecting a parameter $p$, sampling random centre and angle of a single ellipse, determining whether an overlap occurs between the new and existing ellipses, add/reject and repeat until the required number have been collected or reached the maximum number of attempt. At this stage we have a list of $N$ ellipses and compute $Q_{11}$ and $Q_{12}$ from these. Lastly, we render the black and white image and save it alongside a matching label into our dataset}.}
\label{fig: Synthetic Data Generation}
\end{figure}

The simulation uses an algorithm where the rigid ellipses are iteratively introduced into the system and to ensure physical realism, we enforce a non-overlapping constraint. This is achieved through a GPU-accelerated collision detection algorithm (implemented via CUDA) which verifies that the perimeter of a newly proposed ellipse does not intersect with any of the existing particles. If an overlap is detected, the potential new ellipse is discarded and a new position resampled. This process is repeated until the desired number of ellipses is generated. An attempt tracker terminates this algorithm in cases where the density is too high to fit in any new ellipses.

The orientational ordering of the images, which directly relates to the $Q$-tensor, is controlled by a probability parameter $p$ and governs the angular distribution of the ellipses. When $p=0$, the resulting systems resemble the isotropic phase where the molecular orientations are randomly distributed between 0 and $\pi$. As $p$ increases towards $p=1$, the distribution range is progressively restricted, forcing the ellipses to align along a common director. This simulates the ordered nematic phase. The probability parameter controlling the ordering of the training data allows for the generation of a balanced and diverse distribution of sample images and $Q$-tensor values, which are needed for model learning.

Once a valid configuration of non-intersecting ellipses is found, the ground truth $Q$-tensor is computed from the $N$ discrete molecule orientations $\theta_i$. The tensor components $Q_{11}$ and $Q_{12}$ are computed using the discrete $Q$-tensor formulae (\ref{Discrete Q}). The spatial ellipse configurations are converted into a binary image, where ellipses are rendered in white against a black background. This results in a clean, labelled dataset where every synthetically generated image is paired with its exact, mathematically computed $Q$-tensor value.

In order for the distribution of images to be roughly equal (so we avoid over representing isotropic or nematic phases), we exclusively sample over $p = 0,\;0.1,\;0.4,\;0.6,\;1.0$ for both the square and circular images. For each instance of $p$, 10,000 images are generated (totalling 50,000), with each image containing approximately 100 ellipses. Additionally, for reproducibility of the generation process, each generated configuration is assigned a unique integer seed derived from a combination of its associated probability parameter $p$ and its sequential index within the batch. Any specific texture can be reconstructed by referencing the seed embedded in its filename. Testing data can be generated similarly, but with modified seed values, so as not to generate the same training set again. Figure \ref{fig: Synthetic Data Generation} below gives an overview of the data generation process.

Training and test data can be made available upon request.

\end{document}